\documentclass[a4paper,11pt]{article}
\pdfoutput=1 

\usepackage{jheppub} 

\usepackage[T1]{fontenc} 

\usepackage[multiple]{footmisc}
\usepackage{bm}
\usepackage{enumerate}

\title{Multi-fixed point numerical conformal bootstrap: a case study with structured global symmetry}
\author{Matthew T. Dowens}
\author{and Chris A. Hooley}
\affiliation{SUPA, School of Physics and Astronomy, University of St Andrews,\\North Haugh, St Andrews, Fife KY16 9SS, United Kingdom}
\emailAdd{mtd5@st-andrews.ac.uk}
\emailAdd{cah19@st-andrews.ac.uk}
\arxivnumber{2004.14978}

\abstract{In large part, the future utility of modern numerical conformal bootstrap depends on its ability to accurately predict the existence of hitherto unknown non-trivial conformal field theories (CFTs).  Here we investigate the extent to which this is possible in the case where the global symmetry group has a product structure.  We do this by testing for signatures of fixed points using a mixed-correlator bootstrap calculation with a minimal set of input assumptions. This `semi-blind' approach contrasts with other approaches for probing more complicated groups, which `target' known theories with additional spectral assumptions or use the saturation of the single-correlator bootstrap bound as a starting point. As a case study, we select the space of CFTs with product-group symmetry $O(15)\otimes{O}(3)$ in $d=3$ dimensions.  On the assumption that there is only one relevant scalar ($\ell=0$) singlet operator in the theory, we find a single `allowed' region in our chosen space of scaling dimensions.  The scaling dimensions corresponding to two known large-$N$ critical theories, the Heisenberg and the chiral ones, lie on or very near the boundary of this region.  The large-$N$ antichiral point lies well outside the `allowed' region, which is consistent with the expectation that the antichiral theory is unstable, and thus has an additional relevant scalar singlet operator.  We also find a sharp kink in the boundary of the `allowed' region at values of the scaling dimensions that do not correspond to the $(N,M)=(15,3)$ instance of any large-$N$-predicted $O(N) \otimes O(M)$ critical theory.}

\def\be{\begin{equation}}
\def\ee{\end{equation}}
\def\bea{\begin{eqnarray}}
\def\eea{\end{eqnarray}}
\def\co{{\cal O}}
\def\ds{\displaystyle}
\def\ls{\lambda_{\phi\phi\co}^2}
\def\fp{F^{\phi\phi,\phi\phi}_{+, \Delta,\ell}}
\def\fm{F^{\phi\phi,\phi\phi}_{-, \Delta,\ell}}
\def\widthscale{.86}

\begin{document} 
\maketitle
\flushbottom

\section{Introduction}
In a physical system undergoing a continuous phase transition, conformal symmetry emerges when the temperature (or other non-thermal tuning parameter) reaches its critical value.  In the latter half of the twentieth century, it was realized that this dramatic enhancement of the applicable symmetry group opens the door to powerful and unifying theoretical treatments of such systems \cite{cardy1996}.  In particular, the important concept of {\it universality\/} emerged:\ the idea that systems with quite different microscopic physics would be described at criticality by the same conformal field theory (CFT).  As a consequence, the critical exponents describing the behaviour of various physical observables in the approach to criticality would agree exactly.  Indeed, in simple cases these critical exponents are expected to depend only on the dimensionality of space (or, in the case of quantum phase transitions, the effective dimensionality of spacetime) and the group describing the internal symmetry of the theory.

The critical exponents that describe the approach to criticality are related to the scaling dimensions, $\{ \Delta_i \}$, that describe the spatial (or spatiotemporal) correlations in the system when it is precisely at the critical point.  The idea that it might be possible to determine these purely from the requirement of conformal symmetry --- the so-called `conformal bootstrap' approach --- is now several decades old \cite{yellowbook}.  However, it did not become a practical method of placing strong bounds on the scaling dimensions until the recent realization \cite{rattazzi2008,poland2012} that this requirement can be cast in the form of a semidefinite programme, the solvability of which can then be determined computationally.

Since then, notable successes of the conformal bootstrap method include the precise determination of the critical exponents of the 3d Ising model \cite{elshowk2012,elshowk2014,kos2014,kos2016precision}, and of the Heisenberg fixed points that appear in three-dimensional theories with $O(N)$ internal symmetry \cite{kos2014vector,kos2015archipelago,kos2016precision}.  These achievements have been rendered possible by various improvements of the original bootstrap technique, including its mixed-correlator extension \cite{kos2014}, whereby crossing symmetry can now be enforced on a wider range of four-point functions composed of non-identical primary scalar operators transforming in arbitrary representations of the global symmetry group.

\section{Motivation}
Despite the success of numerical conformal bootstrap in placing non-perturbative bounds on the space of critical theories in $d\geq3$, questions remain about its generality and future applicability in extracting information about the space of theories with minimal input. 

Single-correlator bootstrap asks whether there can possibly be a conformal field theory that satisfies crossing symmetry of the four-point correlator of a single scalar field, $\phi$.  It investigates that question for a particular assumed value of the scaling dimension of that field, $\Delta_\phi$, and a particular assumed minimum scaling dimension of operators in another symmetry sector, $\Delta_o$.  These values are then varied to explore the two-dimensional $(\Delta_\phi,\Delta_o)$ plane, and thus that plane is divided into two regions:\ disallowed, and `allowed'.  The quotation marks around `allowed' remind us that any result that a point is `allowed' in fact only means that, to the limited extent investigated by this particular bootstrap calculation, it is not yet provably disallowed.

Such single-correlator calculations have the advantage that the input assumptions are minimal.  However, the resulting `allowed' region is still very large, and real CFTs are only visible if they are extremal solutions lying at its boundary --- in which case they usually appear as sharp `kinks' in the monotonically increasing bound.   It is not clear what determines whether a particular CFT will be visible in this way in a particular $(\Delta_\phi,\Delta_o)$ plane.

In mixed-correlator bootstrap, we instead start with a larger system of mixed four-point correlators, and thus impose crossing symmetry also on four-point functions of non-identical scalar fields belonging to different operator sectors.  For this method to be more constraining than single-correlator bootstrap, we must also impose gaps in operator spectra in these new sectors.  This, however, has the disadvantage of potentially ruling out physical CFTs that don't obey the additional spectral assumptions we have made.

This does not seem to pose a practical difficulty in the case of the space of $d=3$ theories with $\mathbb{Z}_2$ or O($N$) global symmetry, where merely the assumption of a single relevant scalar operator in each of the singlet and vector representations reduces the space of `allowed' scaling dimensions to tiny islands focused around the predicted scaling dimensions of the $d=3$ critical Ising and O($N$) models.
However, as we move to increasingly complicated global symmetry groups, there are likely to be several physical CFTs with disparate operator content, and we are unlikely to see all of them using the same set of spectral assumptions.

There are two possible approaches to this problem.  First, we could specialize our assumptions to `pick out' a particular CFT, and `target' that CFT with one or more mixed-correlator bootstrap calculations.  In this approach, we purposely eliminate CFTs in order to be more constraining.  A notable disadvantage of this approach is that information about the target CFT, obtained either from single-correlator bootstrap calculations or analytical treatments, is needed in order to see its mixed-correlator bootstrap signal.  This is problematic if such information is not available, i.e.\ if we do not know the spectral assumptions to employ to target the CFT --- which, in the case of hitherto unknown CFTs, we will not.  Second, alternatively, we could aim to keep our spectral assumptions as general as we reasonably can, in the hope of seeing signatures of multiple (and perhaps hitherto unknown) CFTs, while still harnessing the extra constraining power of mixed-correlator bootstrap.  The question, then, is whether multiple CFTs will display separate signatures when we perform a mixed-correlator bootstrap calculation with a single set of spectral assumptions.

In this paper we explore this question via a case study:\ the space of $d=3$ CFTs with global symmetry $O(N) \otimes O(M)$. Here, large-$N$ analysis predicts multiple fixed points, a feature which has no analogue in the simpler global symmetry groups mentioned above.  We aim to maintain generality by using minimal spectral assumptions, based on those that were successful in isolating the critical theories in the $d=3$ $\mathbb{Z}_2$ and O($N$) cases.

Beyond the value of the $O(N) \otimes O(M)$ groups as case studies, there are other reasons for being interested in them.  In particular, they are relevant in describing multicritical points in systems with competing ordered phases \cite{jaefari2010,fellows2012,eichhorn2013,hooley2014,eichhorn2014,borchardt2016}; improving our understanding of them could help to shed light on the possible phase diagrams of such systems, which include both the cuprate \cite{lee2006,fradkin2015} and iron-based \cite{dai2015,fernandes2019} families of high-temperature superconductors.  Other methods are of course available for exploring the physics of such multicritical theories, including Monte Carlo treatments \cite{kawamura1992,gezerlis2013} and large-$N$ calculations \cite{pelissetto2001,moshe2003,marino2015}.  However, the conformal bootstrap potentially has advantages over these methods, since (a) unlike Monte Carlo, it exploits from the beginning the fact that the critical theory is conformally invariant, and (b) unlike large-$N$ calculations, it is not dependent on a small-parameter expansion.

There have, to our knowledge, so far been only three applications of the conformal bootstrap method to $O(N) \otimes O(M)$ problems in $d=3$.  In the first \cite{nakayama2014}, by Nakayama and Ohtsuki, the single-correlator bootstrap technique \cite{kos2014vector} is used to explore the space of interacting CFTs in $O(15)\otimes{O}(3)$-symmetric critical theories; in the second \cite{nakayama2015}, by the same authors, a similar analysis is carried out for the $O(3)\otimes O(2)$ and $O(4) \otimes O(2)$ cases.  These two papers impose crossing symmetry on the four-point function of four identical scalar fields transforming in the bifundamental representation of the global symmetry group, i.e.\ as a vector under $O(N)$ and as a vector under $O(M)$, and show that this divides various two-dimensional sections of the space of scaling dimensions into the familiar disallowed and `allowed' regions. For $O(15)\otimes{O}(3)$, they find strong bootstrap evidence of the Heisenberg fixed point lying at the kink in the single-correlator bound. They are unable to isolate the chiral and antichiral fixed points in the same plane of scaling dimensions, since these fixed points lie deep in the `allowed' region. Instead, they look at the space of scaling dimensions in other operator sectors, where they find weaker kinks in single-correlator bounds at locations corresponding to the large-$N$ predictions for the relevant scaling dimensions at the chiral and antichiral fixed points.

The third paper \cite{henriksson2020}, by Henriksson {\it et al.\/}, appeared on the arXiv on the same day as the first version of the work we present here.  It reports the results of both single-correlator and mixed-correlator bootstrap treatments, focussing on the chiral, antichiral, and collinear fixed points of $O(N) \otimes O(M)$ theories for particular choices of $N$ and $M$.  Henriksson {\it et al.\/}\ first reproduce single-correlator bounds in various operator sectors in order to identify the locations of the kinks corresponding to the different fixed points, and then use these locations as input assumptions to mixed-correlator bootstrap calculations.  As a result, they are able to see `allowed islands' corresponding to the chiral, antichiral, and collinear fixed points for various $O(N) \otimes O(M)$ groups.  However, a different and rather specific set of spectral assumptions is needed to see each such island.

For the reasons discussed above, rather than tailoring our analysis to a specific known fixed point in this way, we use just one minimal set of assumptions about the operator spectra of the CFTs.  Furthermore, our assumptions do not rely on prior single-correlator bootstrap calculations.  We look at a large region of the space of scaling dimensions, and hope to see signatures of multiple different critical theories in the results of just one mixed-correlator conformal bootstrap calculation.

The remainder of this paper is structured as follows.  In section \ref{s:analytics} we review some of the known analytical results concerning the scaling dimensions of primary operators in the class of CFTs with $O(N) \otimes O(M)$ global symmetry. In section \ref{s:eqns} we present the operator product expansions (OPEs) necessary to decompose the four-point correlation functions of interest into sums over conformal blocks, and we thus derive the bootstrap equations that encode the crossing symmetry of these correlators.  In section \ref{s:numerics} we describe how these are turned into a semidefinite programme susceptible of numerical treatment.  In section \ref{s:results}, we show the results of our computations:\ as well as strong signatures of the Heisenberg fixed point, and some evidence of signatures of the chiral fixed point, we find a sharp kink on the boundary of the `allowed' region that appears to correspond to a hitherto unknown $O(15) \otimes O(3)$ CFT.  In section \ref{s:discussion}, we discuss the interpretation of our results, and indicate possible lines of future work.

\section{Review of analytical results} \label{s:analytics}
Many analytical results for the scaling dimensions of the primary operators in CFTs with $O(N) \otimes O(M)$ global symmetry are obtained from the renormalisation group (RG) analysis of a Landau-Ginzburg-Wilson Hamiltonian density with the same global symmetry:
\begin{equation}
H = \frac{1}{2}\left(\partial_{\mu}\phi_{a}^{\alpha}\right)\left(\partial_{\mu}\phi_{a}^{\alpha}\right) + \frac{u}{4!}\left(\phi_{a}^{\alpha}\phi_{a}^{\alpha}\right)^{2} + \frac{v}{4!}\left(\phi_{a}^{\alpha}\phi_{b}^{\alpha}\phi_{a}^{\beta}\phi_{b}^{\beta} - \phi_{a}^{\alpha}\phi_{a}^{\alpha}\phi_{b}^{\beta}\phi_{b}^{\beta}\right).
\label{Ham}
\end{equation}
An expansion in $\epsilon\equiv 4-d$, where $d$ is the spatial dimensionality, allows us to compute the beta functions \cite{pelissetto2001}, and thus determine the fixed points and the critical exponents for the case of general $N$ and $M$. Setting the beta functions to zero, we find that there are in general four possible distinct fixed-point solutions:\ the Gaussian fixed point $(u^{*}=v^{*}=0)$; the $O(NM)$-symmetric Heisenberg fixed point $(v^{*}=0)$; and two non-trivial fixed points named chiral and anti-chiral respectively, with non-zero $u^{*}$ and $v^{*}$.

\begin{figure}[tp]
\centering
\includegraphics[width=0.75\textwidth]{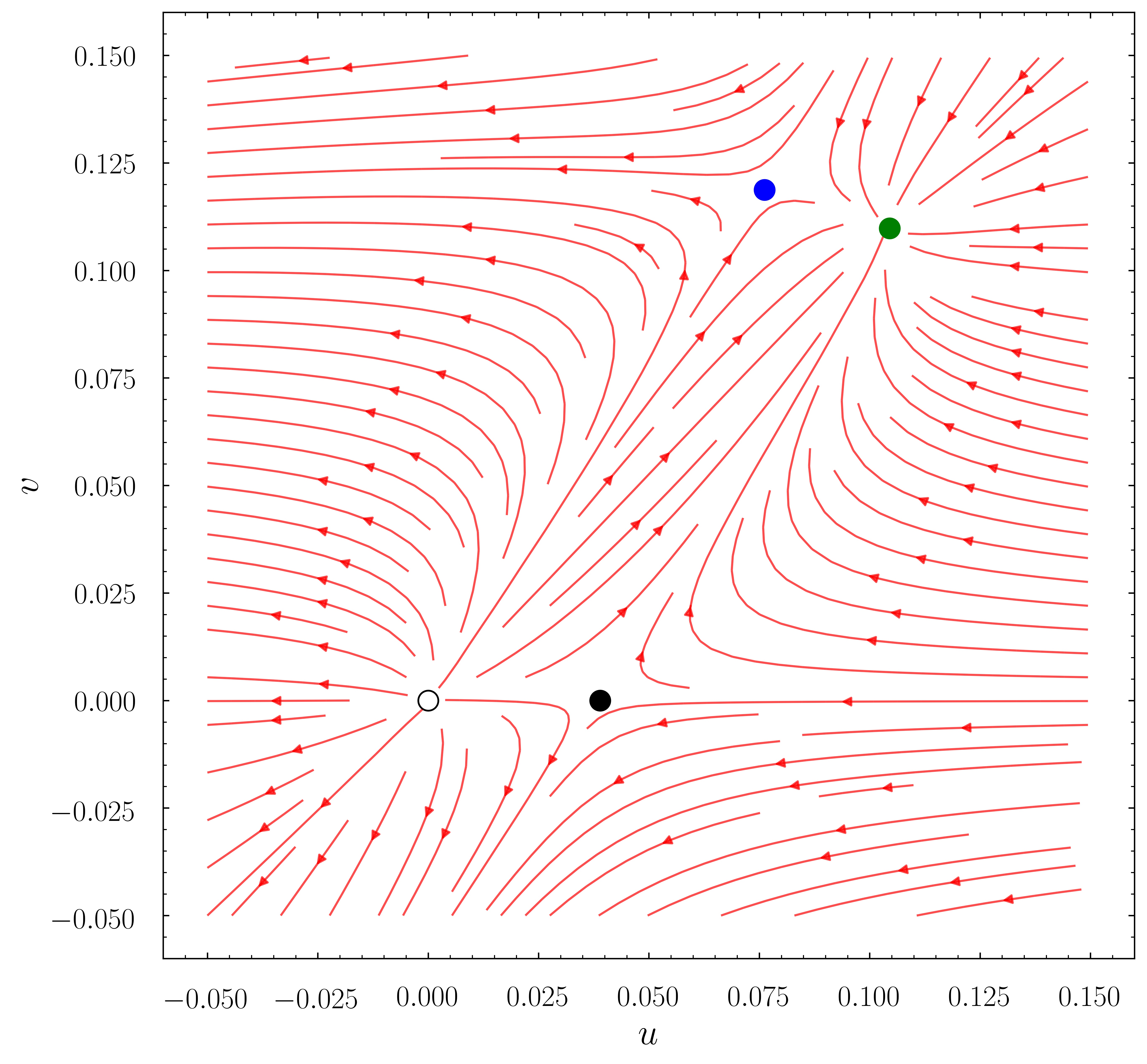}
\caption{\label{fig:0}Renormalisation group flow of the $O(N) \otimes O(M)$-symmetric Hamiltonian (\ref{Ham}) in the $(u,v)$-plane for $(N,M)=(50,3)$. For these values of $N$ and $M$ there are four fixed points:\ the trivial Gaussian fixed point, $(u^{*},v^{*})=(0,0)$ (white); the Heisenberg fixed point, $v^{*}=0$ (black); the chiral fixed point (green); and the anti-chiral fixed point (blue).  This figure is produced using the beta functions calculated via $\epsilon$-expansion in the large-$N$ literature \cite{pelissetto2001} with $\epsilon \equiv 4-d$ set to $1$.  Our choice of $(N,M)$ is the same as that used in Fig.~1 of Nakayama and Ohtsuki's 2014 paper \cite{nakayama2014}.}
\end{figure}

The number of fixed points, and whether they include the chiral and anti-chiral ones, is dependent on the values of $N$ and $M$. There are four regimes.  For a given value of $M$, if $N$ satisfies $N>N^{+}(M)$, we find that all four fixed points are present, with the chiral fixed point the stable one.  An example of the RG flow in this regime is shown in figure \ref{fig:0}.  For the exact form of $N^{+}$, we refer the reader to the original reference. The focus of this paper is $(N,M)=(15,3)$, which places us well within this regime.

We use the $(N,M)=(15,3)$ instances of large-$N$ expressions for the anomalous dimensions found in the literature~\cite{pelissetto2001,moshe2003}. We take the analytical expressions for the critical exponents, $\eta$ and $\nu$, for each of the fixed points in the large-$N$ limit and insert these into expressions for the scaling dimensions of the associated fields: 
\be
\Delta_{\phi}=\frac{1}{2}+\frac{\eta}{2};\hspace{25pt}
\Delta_{s}=3-\frac{1}{\nu}.
\ee
The resulting predictions for the scaling dimensions of $\phi$ and $s$ at the three non-trivial fixed points are
\bea
\mbox{Heisenberg:} & \quad \,\,\,\,\,\,\,\,\big( \Delta_{\phi}^{\text{H}}, \Delta_{s}^{\text{H}} \big) =  (0.5029,1.974); & \\
\mbox{Chiral:} & \quad \,\,\,\,\,\big( \Delta_{\phi}^{\text{C}}, \Delta_{s}^{\text{C}} \big) = (0.518,1.810); & \\
\mbox{Antichiral:} & \,\,\,\,\big( \Delta_{\phi}^{\text{AC}}, \Delta_{s}^{\text{AC}} \big) =  (0.515,1.146). &
\eea
The difference in the number of significant figures is due to the fact that the formula for $\Delta_{\phi}^{\text{H}}$ \cite{moshe2003} includes terms of $O(N^{-3})$, while those for the other scaling dimensions \cite{pelissetto2001} include terms up to $O(N^{-2})$.

We note here a couple of discrepancies between our analytical results and those in previously published works.  First, the expression we obtain for $\Delta_{s}^{\text{C}}$ is
\be
\Delta_{s}^{\rm C} = 2-\frac{16(M+1)}{3\pi^{2}N} - \frac{64 (7M^2+5M-20) + 108(M^2+3M+4) \pi^2}{27\pi^{4}N^{2}},
\ee
which differs from that given in the paper by Henriksson {\it et al.}\ \cite{henriksson2020}:\ they have a factor of $M^2$ in the denominator that we do not.
Second, our computed value of the scaling dimension $\Delta_{s}^{\rm C}=1.810$ disagrees with the value Nakayama and Ohtsuki give \cite{nakayama2014} for their analytical prediction, $\Delta_{s}^{\rm C}=1.796$.  They do not provide a direct reference for their quoted value; however, we note that ours agrees much better than theirs with the value they obtain via spectrum extraction from numerical conformal bootstrap results at the kink in the single-correlator bound that they conjecture corresponds to the chiral point.

\section{Bootstrap equations} \label{s:eqns}
The conformal bootstrap technique starts from one or more four-point correlation functions of the CFT.  Applying the operator product expansion (OPE), each such correlation function can be written as a sum of conformal blocks.  However, since the OPE involves treating the operators in pairs, it can be applied to the four-point correlation function in more than one way, resulting in conformal-block decompositions that look superficially different.  Crossing symmetry is the requirement that the expressions for the four-point correlation function derived by performing the OPE in these different channels should agree with each other.

In the general case where the operators in the four-point correlation function are distinct from each other, it may be decomposed as
\be
\langle p(\vec{x}_1) q(\vec{x}_2) r(\vec{x}_3) t(\vec{x}_4) \rangle = \frac{1}{x_{12}^{\Delta_p + \Delta_q} x_{34}^{\Delta_r + \Delta_t}}
\left( \frac{x_{24}}{x_{14}} \right)^{\Delta_{pq}} \left( \frac{x_{14}}{x_{13}} \right)^{\Delta_{rt}} \sum_{\co} \lambda_{pq\co} \lambda_{rt\co} g^{\Delta_{pq},\Delta_{rt}}_{\Delta,\ell} (u,v). \label{decomp} 
\ee
Here $p$, $q$, $r$, and $t$ are primary operators of the CFT; $\Delta_p$, $\Delta_q$, $\Delta_r$, and $\Delta_t$ are the scaling dimensions of those operators; $\Delta_{ij} \equiv \Delta_i - \Delta_j$; $\vec{x}_{1}$, $\vec{x}_{2}$, $\vec{x}_{3}$, and $\vec{x}_{4}$ are $d$-dimensional position vectors; $x_{ij} \equiv \vert \vec{x}_i - \vec{x}_j \vert$; the sum runs over primary operators $\co$; $\Delta$ and $\ell$ denote respectively the scaling dimension and the spin of $\co$; $\lambda_{pq\co}$ and $\lambda_{rt\co}$ are OPE coefficients; $g^{\Delta_{pq},\Delta_{rt}}_{\Delta,\ell} (u,v)$ is the conformal block associated with the exchange of the operator $\co$; and $u$ and $v$ are the conformal cross-ratios, defined by
\be
u = \frac{x_{12}^2 x_{34}^2}{x_{13}^2 x_{24}^2}; \qquad
v = \frac{x_{14}^2 x_{23}^2}{x_{13}^2 x_{24}^2}. \label{crossratios}
\ee
Aside from small notational changes, (\ref{decomp}) is just equation (2.1) of ref.~\cite{kos2015archipelago}.

In a CFT with an internal symmetry group, the primary operators may be classified according to their transformation properties under that group.  For the direct-product group $O(N) \otimes O(M)$, on which we focus in this work, we label the relevant representations $XY$, where $X,Y \in \{ S,V,T,A \}$.  The letters in this list stand respectively for singlet, vector, traceless symmetric tensor, and antisymmetric tensor.  $X$ and $Y$ respectively denote the transformation properties of the operator under $O(N)$ and $O(M)$.

For a given choice of representation for each of the external operators $p$, $q$, $r$, and $t$, we can use the fusion rules of the group to determine the allowed representations of the exchanged operator $\co$.  For $O(N) \otimes O(M)$, the fusion rules that we shall need in this work are
\bea
s \times s & \sim & \sum_{SS^{+}} \co; \label{ope1} \\
\phi_{i\alpha} \times s & \sim & \sum_{VV^{\pm}} \co_{i\alpha}; \label{ope2} \\
\phi_{i\alpha} \times \phi_{j\beta} & \sim & 
\sum_{SS^{+}} \delta_{ij} \delta_{\alpha\beta} \co
+ \sum_{ST^{+}} \delta_{ij} \co_{(\alpha\beta)} 
+ \sum_{SA^{-}} \delta_{ij} \co_{[\alpha\beta]} \nonumber \\ & & \quad
+ \sum_{TS^{+}} \delta_{\alpha\beta} \co_{(ij)}
+ \sum_{TT^{+}} \co_{(ij)(\alpha\beta)}
+ \sum_{TA^{-}} \co_{(ij)[\alpha\beta]} \nonumber \\ & & 
+ \sum_{AS^{-}} \delta_{\alpha\beta} \co_{[ij]}
+ \sum_{AT^{-}} \co_{[ij](\alpha\beta)}
+ \sum_{AA^{+}} \co_{[ij][\alpha\beta]}. \label{ope3}
\eea
Here $s$ denotes the most relevant primary operator in the $\{SS\,|\,\ell=0\}$ sector, and $\phi_{i\alpha}$ the most relevant primary operator in the $\{VV\,|\,\ell=0\}$ sector.  Roman indices are associated with the $O(N)$ subgroup and thus take values from $1$ to $N$; Greek indices are associated with the $O(M)$ subgroup and thus take values from $1$ to $M$.  The symbol $\co$ denotes a primary operator belonging to the sector indicated below the relevant summation sign.  Parentheses $(\alpha\beta)$ denote symmetrisation, while brackets $[\alpha\beta]$ denote antisymmetrisation.  A ${+}$ or a ${-}$ superscript indicates that the sum in question runs over just the even- or just the odd-spin primary fields in that representation:\ such restrictions arise from the requirement that the fusion rule be symmetric under the interchange of the bosonic operators on the left-hand side.  No such restriction applies in (\ref{ope2}), since in this case the fused operators $\phi_{i\alpha}$ and $s$ are distinguishable. The sum therefore runs over all spins, and we indicate this with a $\pm$ sign.

To derive our crossing symmetry equations, we consider four different four-point correlation functions:
\bea
G^{(VV)^4}_{ijkl\alpha\beta\gamma\delta} & \equiv & \langle \phi_{i\alpha}(\vec{x}_1) \phi_{j\beta}(\vec{x}_2) \phi_{k\gamma}(\vec{x}_3) \phi_{l\delta}(\vec{x}_4) \rangle; \label{vv4} \\
G^{(SS)^4} & \equiv & \langle s(\vec{x}_1) s(\vec{x}_2) s(\vec{x}_3) s(\vec{x}_4) \rangle; \label{ss4} \\
G^{(VV)^2 (SS)^2}_{ij\alpha\beta} & \equiv & \langle \phi_{i\alpha}(\vec{x}_1) \phi_{j\beta}(\vec{x}_2) s(\vec{x}_3) s(\vec{x}_4) \rangle; \label{vv2ss2} \\
G^{(VV)(SS)(VV)(SS)}_{ij\alpha\beta} & \equiv & \langle \phi_{i\alpha}(\vec{x}_1) s(\vec{x}_2) \phi_{j\beta}(\vec{x}_3) s(\vec{x}_4) \rangle.\label{vvssvvss}
\eea
For each of these correlators, we equate the results of two different conformal block decompositions of the correlator:\ one where the first operator is paired with the second, as in (\ref{decomp}), and one where the first operator is paired with the fourth.  In practice, the latter decomposition is obtained simply by making an exchange of labels such as $q \leftrightarrow t$ and $\vec{x}_2 \leftrightarrow \vec{x}_4$ in (\ref{decomp}).  After separating the coefficients of different fundamental tensor structures, we obtain a total of thirteen bootstrap equations:\ nine from (\ref{vv4}), one from (\ref{ss4}), two from (\ref{vv2ss2}), and one from (\ref{vvssvvss}).  These constraints can be encoded in a single 13-dimensional vectorial sum rule,
\bea
0 = \sum_{{SS}^{+}}
\begin{pmatrix}
\lambda_{\phi\phi\co} & \lambda_{ss\co}
\end{pmatrix}
\vec{V}_{SS,\Delta, \ell}
\begin{pmatrix}
\lambda_{\phi\phi\co} \\
\lambda_{ss\co}
\end{pmatrix}
 + \sum_{{ST}^{+}}\lambda_{\phi\phi\co}^{2}\vec{V}_{ST,\Delta, \ell} + \sum_{{SA}^{-}}\lambda_{\phi\phi\co}^{2}\vec{V}_{SA,\Delta, \ell} \nonumber \\
+ \sum_{{TS}^{+}}\lambda_{\phi\phi\co}^{2}\vec{V}_{TS,\Delta, \ell} + \sum_{{TT}^{+}}\lambda_{\phi\phi\co}^{2}\vec{V}_{TT,\Delta, \ell} + \sum_{{TA}^{-}}\lambda_{\phi\phi\co}^{2}\vec{V}_{TA,\Delta, \ell} \nonumber \\
+ \sum_{{AS}^{-}}\lambda_{\phi\phi\co}^{2}\vec{V}_{AS,\Delta, \ell} + \sum_{{AT}^{-}}\lambda_{\phi\phi\co}^{2}\vec{V}_{AT,\Delta, \ell} + \sum_{{AA}^{+}}\lambda_{\phi\phi\co}^{2}\vec{V}_{AA,\Delta, \ell} \nonumber \\
+ \sum_{{VV}^{+}}\lambda_{\phi{s}\co}^{2}\vec{V}_{VV+,\Delta, \ell} + \sum_{{VV}^{-}}\lambda_{\phi{s}\co}^{2}\vec{V}_{VV-,\Delta, \ell}. \label{sum_rule}
\eea
Here, $\vec{V}_{SS,\Delta,\ell}$ is a 13-vector of $2 \times 2$ matrices and all the other ${\vec V}_{XY,\Delta,\ell}$ are 13-vectors of $1 \times 1$ matrices, i.e.\ scalars.  All of these vectors are composed of various combinations of the convolved conformal blocks,
\be
F_{\pm,\Delta,\ell}^{pq,rt}(u,v) \equiv v^{\frac{\Delta_{r}+\Delta_{q}}{2}}g_{\Delta,\ell}^{\Delta_{pq},\Delta_{rt}}(u,v) \pm u^{\frac{\Delta_{r}+\Delta_{q}}{2}}g_{\Delta,\ell}^{\Delta_{pq},\Delta_{rt}}(v,u).
\ee
A detailed derivation, including the explicit form of the vectors in terms of the convolved conformal blocks, is provided in appendix \ref{app:bootstrapeqns}. 

\section{Computational solution} \label{s:numerics}
For a given set of assumptions about the spectrum of scaling dimensions in each operator sector, our vectorial sum rule (\ref{sum_rule}) provides an associated set of constraints on the OPE coefficients $\lambda_{pq\co}$ that appear in the decompositions (\ref{decomp}).  These constraints may be mutually inconsistent:\ if they are, then our assumptions about the spectrum are inconsistent with crossing symmetry, and thus cannot be obeyed by any CFT.

To determine whether the constraints (\ref{sum_rule}) are mutually inconsistent, we search for a linear transformation under which the right-hand side of the sum rule becomes positive definite for any choice of the values of the OPE coefficients consistent with unitarity.  If such a transformation can be found, then the right-hand side of the transformed version of (\ref{sum_rule}) is strictly positive, while the left-hand side is zero.  This is a contradiction, and thus shows that our assumptions on the operator spectrum cannot have been correct.  By this means, a particular subspace of the space of scaling dimensions can be ruled out.

In practice, we search for a vector of linear functionals, $\vec{y} = \left(y_{1}, y_{2}, ..., y_{13}\right)$, each of which maps the convolved conformal blocks to a linear combination of a finite number of their derivatives at the crossing-symmetric point, $(u,v) = \left(u_{*},v_{*}\right) \equiv \left(\frac{1}{4},\frac{1}{4}\right)$:
\be
y_{k}\left(F_{\pm,\Delta,\ell}^{pq,rt}(a,b)\right) = \sum_{(m,n) \in {\cal D}} y_{kmn} \left. \partial_{a}^{m}\partial_{b}^{n}F_{\pm,\Delta,\ell}^{pq,rt}(a,b)\right\vert_{a=1,b=0}.
\label{taylor_expansion}
\ee
Note that we have switched from the $(u,v)$ to the $(a,b)$ coordinate system, in which the crossing symmetric point is at $(a,b) = (a_{*},b_{*}) \equiv (1,0)$ \cite{elshowk2012,behan2017}.  The set of derivatives ${\cal D}$ contains all pairs $(m,n)$ that satisfy the following conditions \cite{hogervorst2013diagonal,behan2017}:\footnote{This is the same set of points $(m,n)$ as used in \cite{behan2017}.  However, equation (2.20b) of that paper contains a typographical error:\ their $n-n_{\rm max}$ should read $n_{\rm max}-n$, as we have here.}
\bea
n & \in & \left\{ 0, \ldots, n_{\rm max} \right\}; \\
m & \in & \left\{ 0, \ldots, 2(n_{\rm max}-n)+m_{\rm max} \right\}.
\eea
This set is parametrised by two independent integers, $m_{\rm max}$ and $n_{\rm max}$.  For all results presented in this paper, $m_{\rm max} = n_{\rm max} - 2$; therefore we specify only $n_{\rm max}$ in what follows.

To allow operators of arbitrary scaling dimension in one or more sectors of the spectrum, we must replace the convolved conformal blocks that appear in (\ref{taylor_expansion}) by rational approximations to them, which we determine via appropriate recursion relations \cite{kos2014vector,elshowk2014}.  In this work we truncate the polynomial order of these conformal block approximations at $k_{\rm max} = 40$. We also restrict the value of the spin $\ell$ to lie within the range $0 \leqslant \ell \leqslant \ell_{\rm max}$ where $\ell_{\rm max} = 23$.

The assumptions about the spectrum of the CFT that we aim to test are as follows:
\begin{enumerate}
	\item The $d=3$ CFT has global symmetry $O(15) \otimes O(3)$ and is unitary.
	\item The spectrum in the $\{ VV{+}\,\vert\,\ell=0 \}$ sector contains only one relevant operator, the scaling dimension of which is $\Delta_{\phi}$.  All other operators in this sector have scaling dimensions $\Delta > d$.
	\item The spectrum in the $\{ SS\,\vert\,\ell=0 \}$ sector contains only one relevant operator, the scaling dimension of which is $\Delta_{s}$.  All other operators in this sector have scaling dimensions $\Delta > d$.
	\item The OPE relation $\lambda_{\phi\phi s}=\lambda_{\phi s\phi}$ holds.
\end{enumerate}
This implies that the inequalities that the functionals $\vec{y}$ must satisfy if they are to prove the crossing-symmetry sum rule (\ref{sum_rule}) inconsistent are as follows:
\be
\begin{array}{rclcl}
\ds \begin{pmatrix}
1 & & 1
\end{pmatrix}
\vec{y}\cdot\vec{V}_{SS,0,0}
\begin{pmatrix}
1 \\ 1
\end{pmatrix}
& > & 0; & \qquad & \mbox{\small [\,unit operator / normalisation\,]} \vspace{2mm} \\ \vspace{2mm}
\vec{y}\cdot\vec{V}_{XY,\Delta,\ell} & \geqslant & 0; & & \Delta\geqslant\Delta_{\text{unitarity}}(\ell) \\ \vspace{2mm}
\vec{y}\cdot\vec{V}_{VV+,\Delta,\ell} & \geqslant & 0; & & \Delta\geqslant \Delta_{\text{unitarity}}(\ell), \,\,\ell>0 \\ \vspace{2mm}
\vec{y}\cdot\vec{V}_{VV+,\Delta,0} & \geqslant & 0; & & \Delta\geqslant d \\ \vspace{2mm}
\vec{y}\cdot\vec{V}_{SS,\Delta,\ell} & \succeq & 0; & & \Delta\geqslant \Delta_{\text{unitarity}}(\ell), \,\,\ell>0 \\ \vspace{2mm}
\vec{y}\cdot\vec{V}_{SS,\Delta,0} & \succeq & 0; & & \Delta\geqslant d \\ \vspace{2mm}
\vec{y}\cdot\left(\vec{V}_{SS,\Delta_{s},0}+\vec{V}_{VV+,\Delta_{\phi},0}\otimes
\begin{pmatrix}
1 & 0 \\
0 & 0
\end{pmatrix}\right) & \succeq & 0. & & 
\end{array}
\label{sdp}
\ee
Here $d=3$ is the dimensionality of space, while the unitarity bound for the scaling dimensions is given by
\be
\Delta_{\text{unitarity}}(\ell) = \left\{ \begin{array}{lll}
\ds \frac{d-2}{2} & & \ell=0; \\
& & \\
\ell+d-2 & \qquad & \ell>0.
\end{array} \right.
\ee
The second inequality in (\ref{sdp}) actually represents a set of nine inequalities, one for each value of $XY \in \left\{ ST, SA, TS, TT, TA, AS, AT, AA, VV{-} \right\}$.  In each case the spin $\ell$ runs over either the even integers with $0 \leqslant \ell \leqslant \ell_{\rm max}$ or the odd integers with $0 \leqslant \ell \leqslant \ell_{\rm max}$, depending on the sign attached to that representation in (\ref{sum_rule}).

If, for a given choice of $\Delta_\phi$ and $\Delta_s$, we find a $\vec{y}$ satisfying (\ref{sdp}), then the programme is said to be dual feasible and we rule out that particular set of assumptions about the spectrum.  We iteratively revise our assumptions by changing the values of $\Delta_{\phi}$ and $\Delta_{s}$, thus modifying the semidefinite programme, and test dual feasibility at each such point.  By thus ruling out certain possible ranges of scaling dimensions, we place bounds on the scaling dimensions of these relevant operators.  The resulting grids of `allowed' and disallowed scaling dimensions form our main results and are presented and discussed in the next section.

Our implementation uses a modified version of PyCFTBoot \cite{behan2017} as a front-end for generating semidefinite programmes from our bootstrap equations. These are then solved by SDPB, the arbitrary-precision semidefinite programme solver designed for conformal bootstrap calculations \cite{simmons2015, landry2019}. We run SDPB with {\tt precision=1024}, {\tt findPrimalFeasible=true}, {\tt findDualFeasible=true}, {\tt primalErrorThreshold=$\tt 10^{\tt -30}$}, and {\tt dualErrorThreshold=$\tt 10^{\tt -15}$}, leaving all other solver settings at their default values.  Usually one of two things happens:\ either a primal feasible solution is returned relatively quickly, in which case we say that the point is `allowed'; or a dual feasible solution is eventually found, in which case we say that it is disallowed.  The quotation marks around `allowed' are deliberate:\ what this outcome really means is just that the point $(\Delta_\phi,\Delta_s)$ is not ruled out by crossing symmetry constraints at our chosen derivative order $n_{\rm max}$. For full details on the software and underlying algorithms, we refer the reader to the original references.

\begin{figure}[tp]
\centering
\includegraphics[width=\widthscale\textwidth]{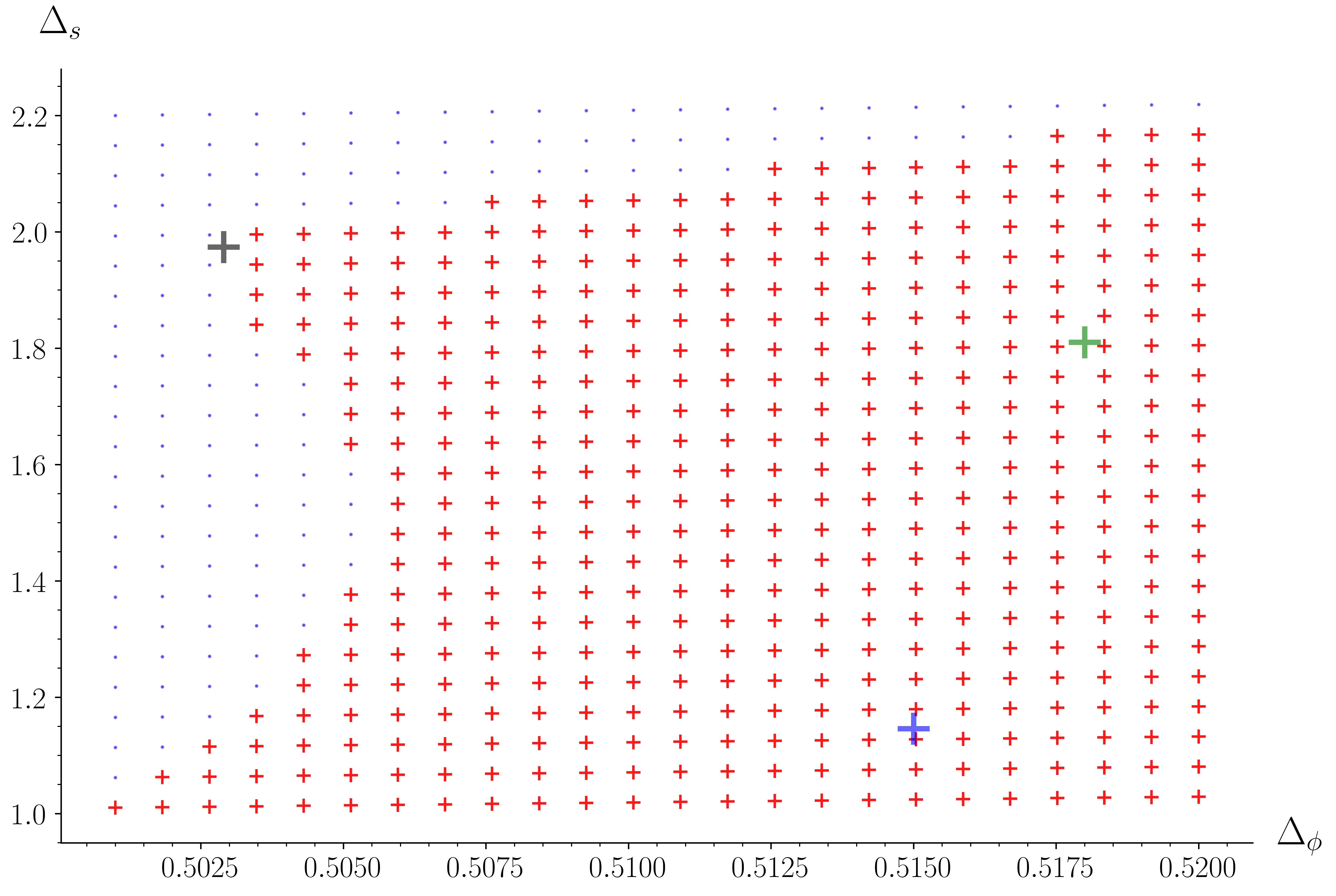}
\caption{\label{fig:1} Mixed-correlator conformal bootstrap results for derivative order $n_{\rm max} = 5$, under the assumptions of (i) precisely one relevant $\ell=0$ singlet ($SS$) operator with scaling dimension $\Delta_{s}$, and (ii) precisely one relevant $\ell=0$ bifundamental ($VV$) operator with scaling dimension $\Delta_{\phi}$.  Red crosses mark the points that were determined to be primal feasible, i.e.\ `allowed'; blue dots mark the points that were determined to be dual feasible, i.e.\ disallowed.  The large crosses mark the large-$N$ predictions for the scaling dimensions of the Heisenberg point (black), the chiral point (green), and the antichiral point (blue).}
\end{figure}

\section{Results} \label{s:results}
Our first set of results, shown in figure \ref{fig:1}, are obtained at the relatively low derivative order $n_{\rm max}=5$.  The points that were determined to be primal feasible, i.e.\ `allowed', are marked with red crosses; the ones that were determined to be dual feasible, i.e.\ disallowed, are marked with blue dots.  It is worth comparing these results to the single-correlator bootstrap results shown in figure 2 of Nakayama and Ohtsuki's 2014 paper \cite{nakayama2014}.  There is not that much difference, except that in our mixed-correlator results we have ruled out an extra set of points in the region $\Delta_\phi \lesssim 0.5055$.  As a result of this, the left-hand boundary of the `allowed' region is now concave.

\begin{figure}[tp]
\centering
\includegraphics[width=\widthscale\textwidth]{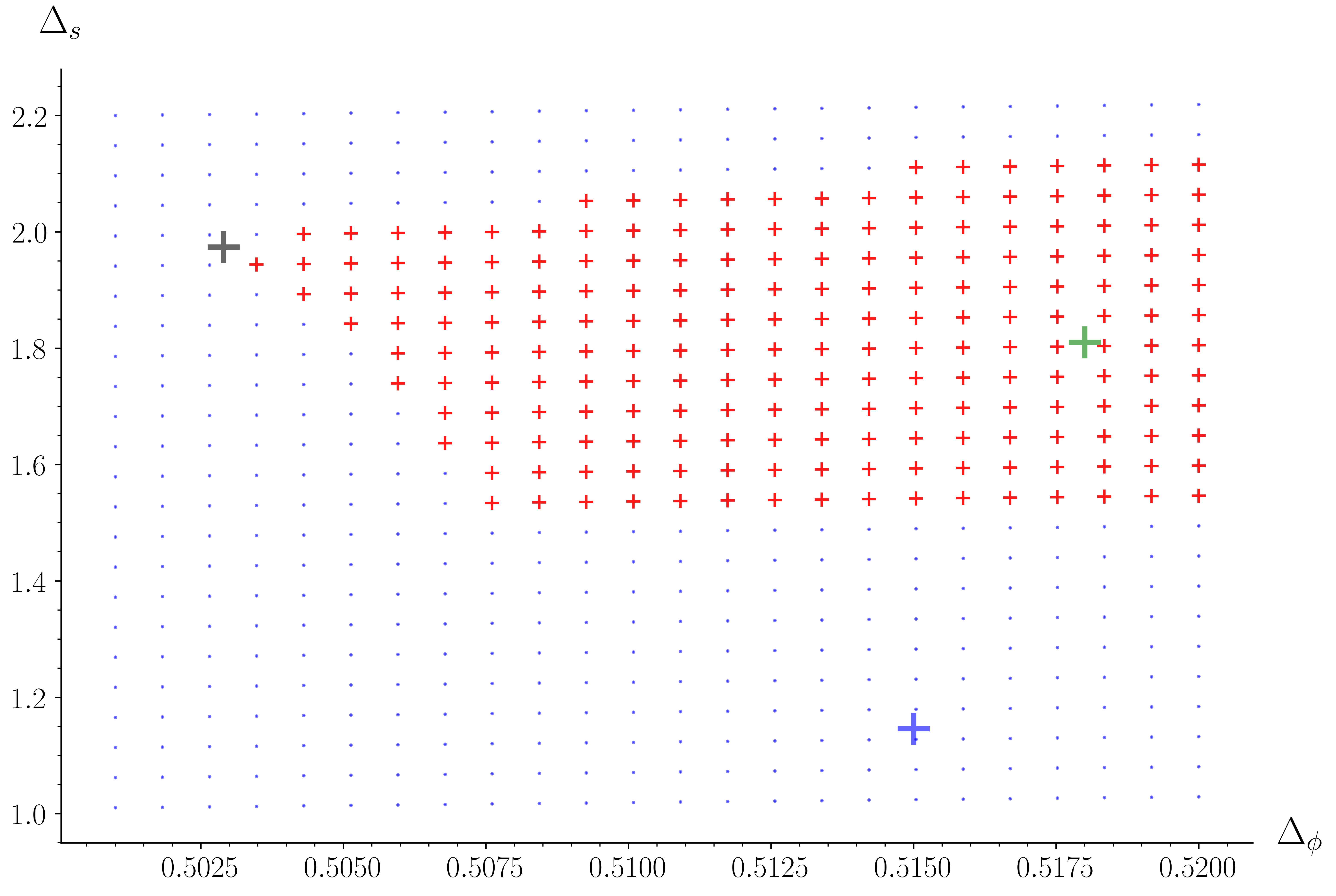}
\caption{\label{fig:2} Mixed-correlator conformal bootstrap results for derivative order $n_{\rm max} = 6$.  The assumptions are the same as in figure \ref{fig:1}, except that we have not imposed the OPE relation, i.e.\ we have omitted the final inequality in (\ref{sdp}).  This does not materially affect the extent or shape of the `allowed' region.  Red crosses mark the points that were determined to be primal feasible, i.e.\ `allowed'; blue dots mark the points that were determined to be dual feasible, i.e.\ disallowed.  The large crosses mark the large-$N$ predictions for the scaling dimensions of the Heisenberg point (black), the chiral point (green), and the antichiral point (blue).  Notice that the antichiral point is now deep in the disallowed region:\ see text for discussion.}
\end{figure}

For comparison, we show in figure \ref{fig:2} the results for derivative order $n_{\rm max} = 6$.  (These results have been calculated without imposing the OPE relation, i.e.\ omitting the final inequality in (\ref{sdp}), but this does not materially affect the extent or shape of the `allowed' region.)  The `allowed' region has been significantly reduced; as a result, the antichiral point is now deep in the disallowed region.  This does not, of course, mean that there is no antichiral CFT for $O(15) \otimes O(3)$; it just means that the antichiral CFT violates the set of assumptions under which we set up the semidefinite programme.  Presumably in this case the invalid assumption is that there is only one relevant operator in the $\left\{ SS\,\vert\,\ell=0 \right\}$ sector.  The antichiral point is unstable --- see figure 1 of Nakayama and Ohtsuki's 2014 paper \cite{nakayama2014} --- and thus its CFT should have a second relevant scalar $SS$ operator, called $S'$ in the large-$N$ literature.

\begin{figure}[tp]
\centering
\includegraphics[width=\widthscale\textwidth]{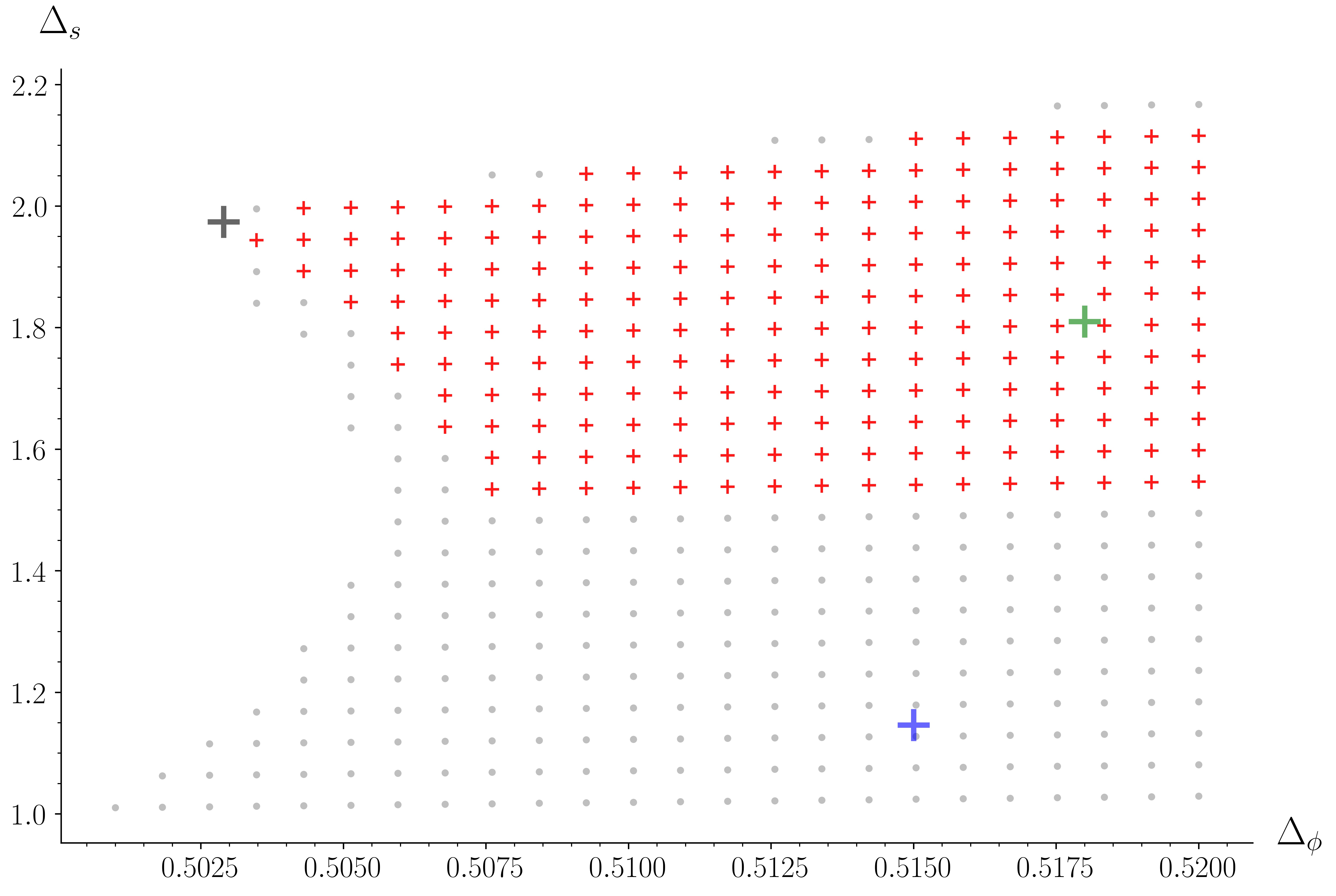}
\caption{\label{fig:3} A comparison of our mixed-correlator conformal bootstrap results between derivative orders $n_{\rm max} = 5$ and $n_{\rm max}=6$.  Grey dots mark the points that were determined to be primal feasible, i.e.\ `allowed', at derivative order $n_{\rm max}=5$; red crosses mark the subset of these that were determined also to be primal feasible, i.e.\ `alllowed', at derivative order $n_{\rm max}=6$.  The large crosses mark the large-$N$ predictions for the scaling dimensions of the Heisenberg point (black), the chiral point (green), and the antichiral point (blue).}
\end{figure}

Figure \ref{fig:3} shows an overlay comparison between the $n_{\rm max}=5$ and $n_{\rm max}=6$ data sets.  Here we have not indicated the disallowed points at all.  Grey dots indicate the points that are `allowed' at derivative order $n_{\rm max}=5$; red crosses indicate the subset of these that remain `allowed' at derivative order $n_{\rm max}=6$.  This figure illustrates that, as well as a drastic reduction in the size of the `allowed' region from below, a few points above and to the left of it have also been ruled out between $n_{\rm max}=5$ and $n_{\rm max}=6$.

\begin{figure}[tp]
\centering
\includegraphics[width=\widthscale\textwidth]{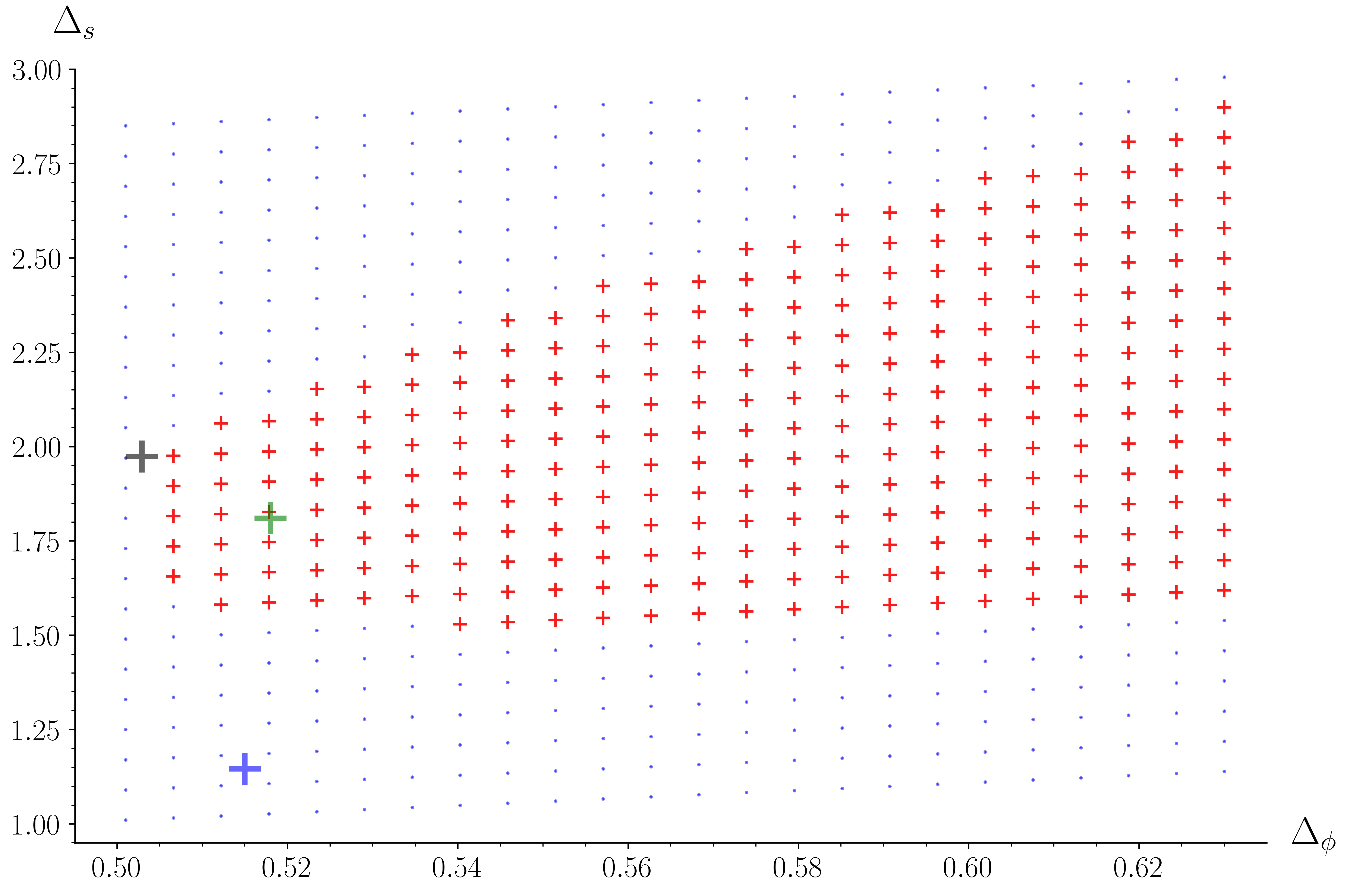}
\caption{\label{fig:4} Mixed-correlator conformal bootstrap results for derivative order $n_{\rm max} = 6$, under the same assumptions as in figure \ref{fig:2} but for a larger range of scaling dimensions.  Red crosses mark the points that were determined to be primal feasible, i.e.\ `allowed'; blue dots mark the points that were determined to be dual feasible, i.e.\ disallowed.  The large crosses mark the large-$N$ predictions for the scaling dimensions of the Heisenberg point (black), the chiral point (green), and the antichiral point (blue).}
\end{figure}

In figure \ref{fig:4} we present another $n_{\rm max}=6$ data set, this time for a wider field of view.  This is simply to demonstrate that there are no further sharp features in the boundary of the `allowed' region.  The apparent kink in the lower boundary at $\Delta_\phi \approx 0.54$ is just the effect of our finite resolution:\ the gradient of the lower boundary of the `allowed' region is slightly less than 1, while the gradient of the rows of our sampling grid is precisely 1.  (This is because the conformal blocks depend only on the difference between $\Delta_\phi$ and $\Delta_s$; thus, as pointed out in \cite{kos2014}, it is computationally more efficient to sample along lines where this difference remains constant.)

\begin{figure}[tp]
\centering
\includegraphics[width=\widthscale\textwidth]{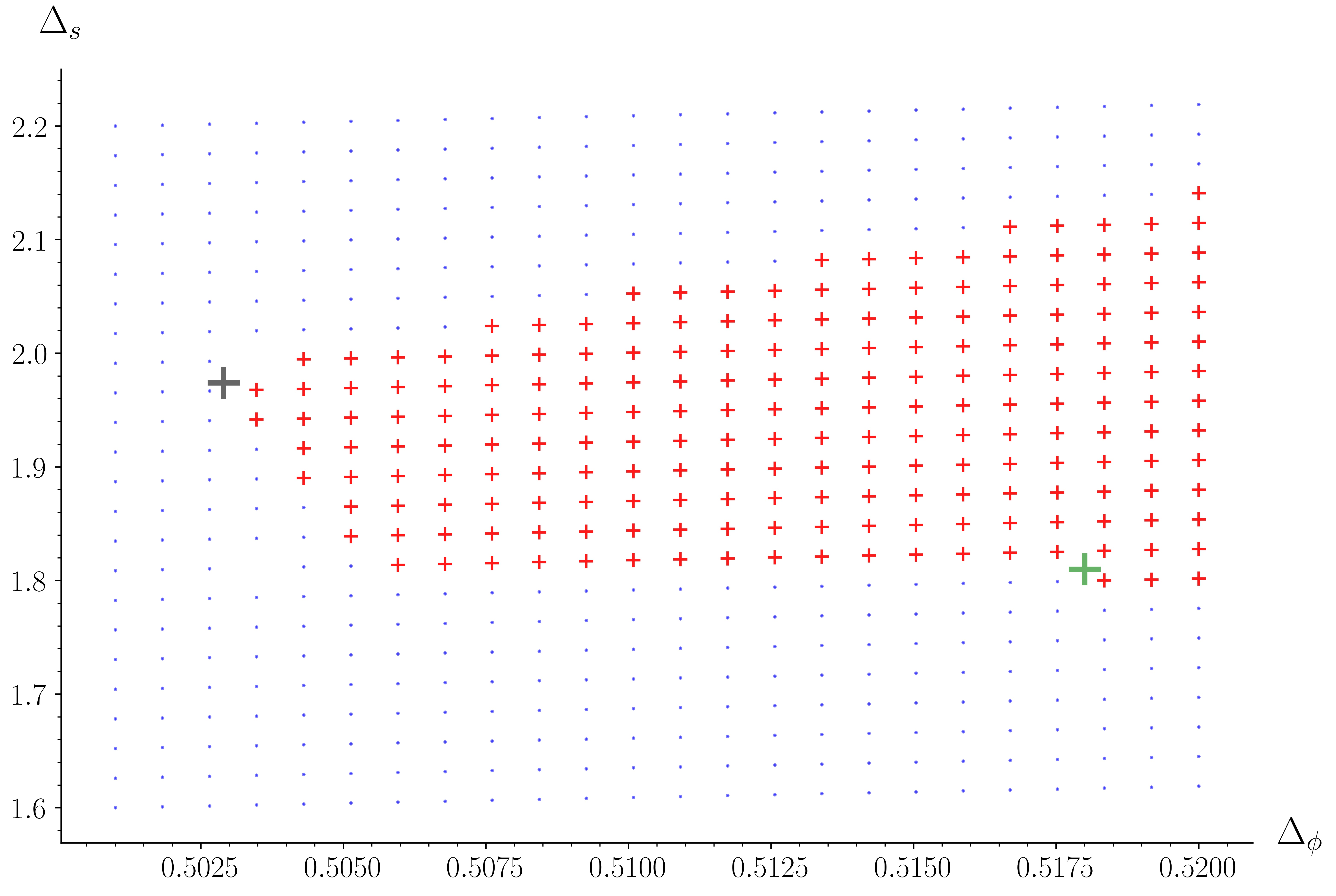}
\caption{\label{fig:5} Mixed-correlator conformal bootstrap results for derivative order $n_{\rm max} = 8$, under the same assumptions as in figure \ref{fig:1}.  Red crosses mark the points that were determined to be primal feasible, i.e.\ `allowed'; blue dots mark the points that were determined to be dual feasible, i.e.\ disallowed.  The large crosses mark the large-$N$ predictions for the scaling dimensions of the Heisenberg point (black) and the chiral point (green).  Note that, at this resolution, both points appear to be on the boundary of the `allowed' region.  Note also the sharp kink at $(\Delta_\phi,\Delta_s) \approx (0.506,1.81)$, which does not correspond to the $(N,M)=(15,3)$ instance of any large-$N$-predicted $O(N) \otimes O(M)$ critical theory.}
\end{figure}

In figure \ref{fig:5}, we increase the derivative order to $n_{\rm max} = 8$ and focus on the tip of the `allowed' region. The increase in derivative order from $n_{\rm max} = 6$ has the effect of cutting away at the bottom of the `allowed' region. At this resolution, the resulting `allowed' region contains both the Heisenberg and chiral fixed points, which both appear to lie near its edge. We also observe the appearance of a seemingly sharp kink in the boundary at $\left(\Delta_{\phi}, \Delta_{s}\right)\approx\left(0.506, 1.81\right)$.

\begin{figure}[tp]
\centering
\includegraphics[width=\widthscale\textwidth]{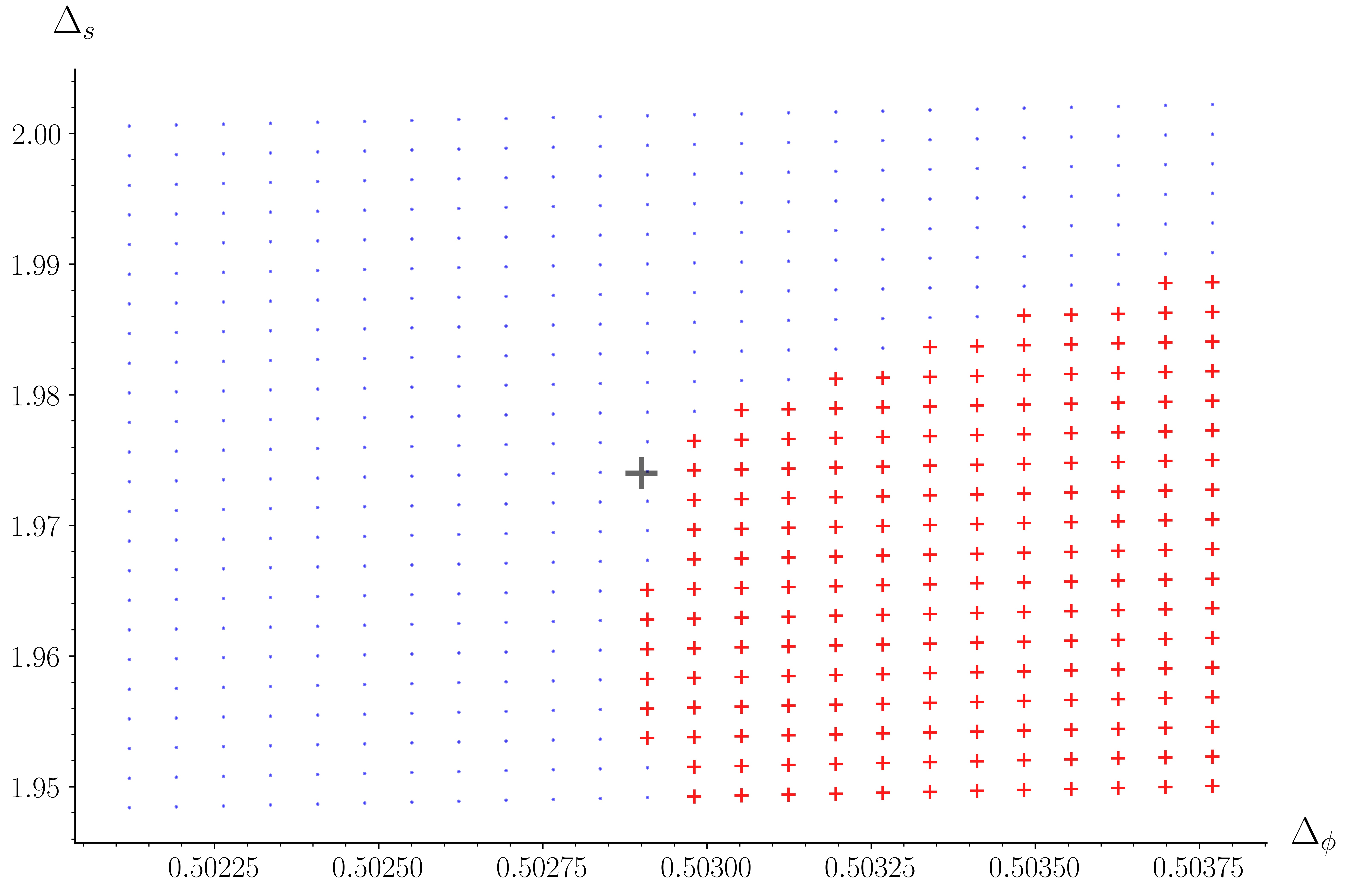}
\caption{\label{fig:6} Mixed-correlator conformal bootstrap results for derivative order $n_{\rm max} = 8$, under the same assumptions as in figure \ref{fig:1}, for a small region around the large-$N$ Heisenberg point predictions.  Red crosses mark the points that were determined to be primal feasible, i.e.\ `allowed'; blue dots mark the points that were determined to be dual feasible, i.e.\ disallowed.  The large black cross marks the large-$N$ prediction for the scaling dimensions of the Heisenberg point.}
\end{figure}

\begin{figure}[tp]
\centering
\includegraphics[width=\widthscale\textwidth]{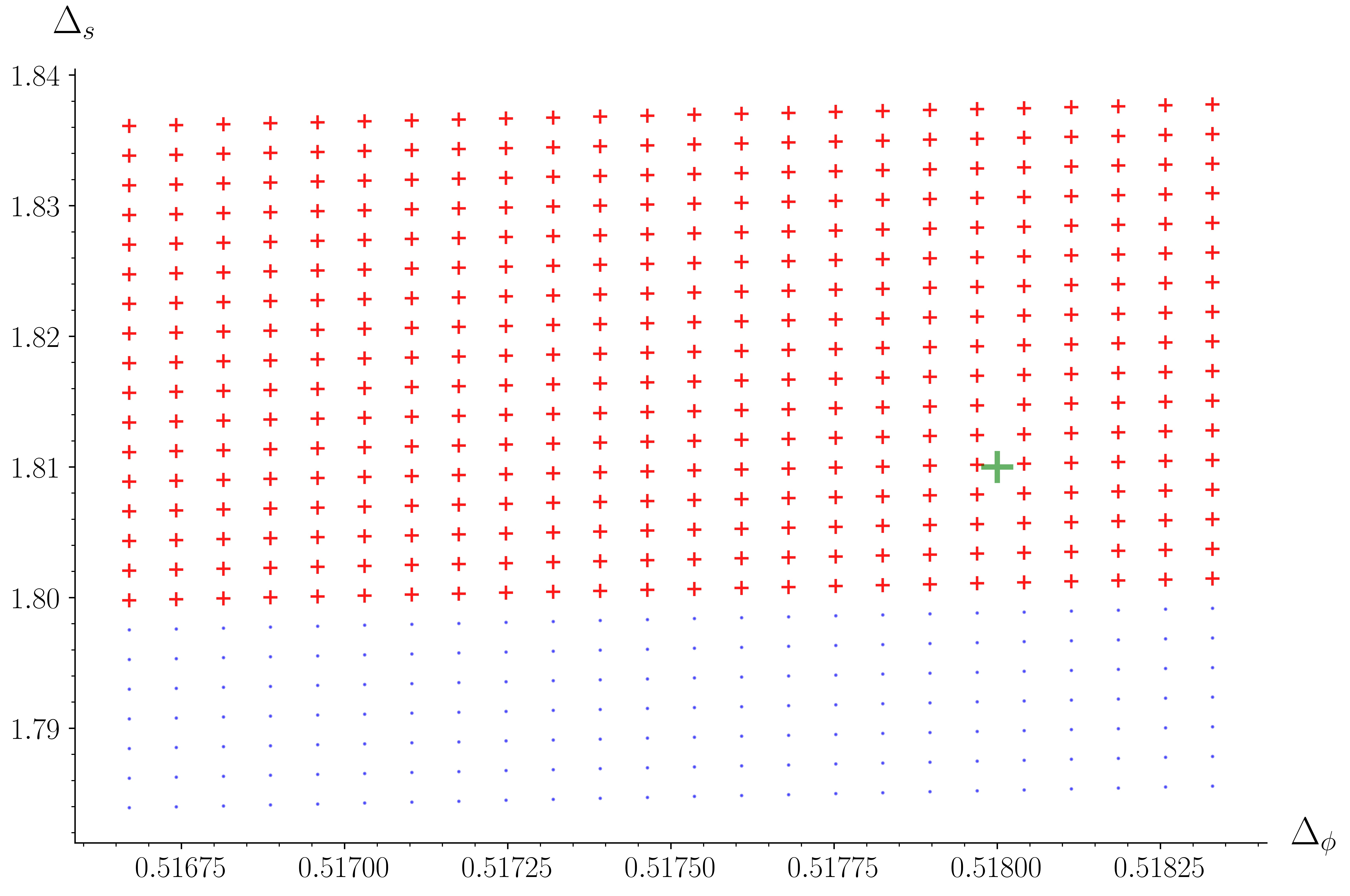}
\caption{\label{fig:7} Mixed-correlator conformal bootstrap results for derivative order $n_{\rm max} = 8$, under the same assumptions as in figure \ref{fig:1}, for a small region around the large-$N$ chiral point predictions.  Red crosses mark the points that were determined to be primal feasible, i.e.\ `allowed'; blue dots mark the points that were determined to be dual feasible, i.e.\ disallowed.  The large green cross marks the large-$N$ prediction for the scaling dimensions of the chiral point.}
\end{figure}

Figure \ref{fig:6} shows a close-up view of the region around the large-$N$-predicted location of the Heisenberg fixed point.  At this resolution, the mixed-correlator bound signals the fixed point with a sharp kink and the large-$N$ prediction lies very close to the edge of the `allowed' region. Figure \ref{fig:7} shows a similar close-up of the region around the large-$N$-predicted location of the chiral fixed point. This reveals that, given our spectral assumptions, this location is `allowed' by crossing symmetry at this derivative order, and indeed lies within the `allowed' region rather than on its boundary.  We note that higher-order corrections in the $1/N$-expansion would change the location of the green cross, but it seems unlikely that they would move it all the way to the boundary shown here.

\begin{figure}[tp]
\centering
\includegraphics[width=\widthscale\textwidth]{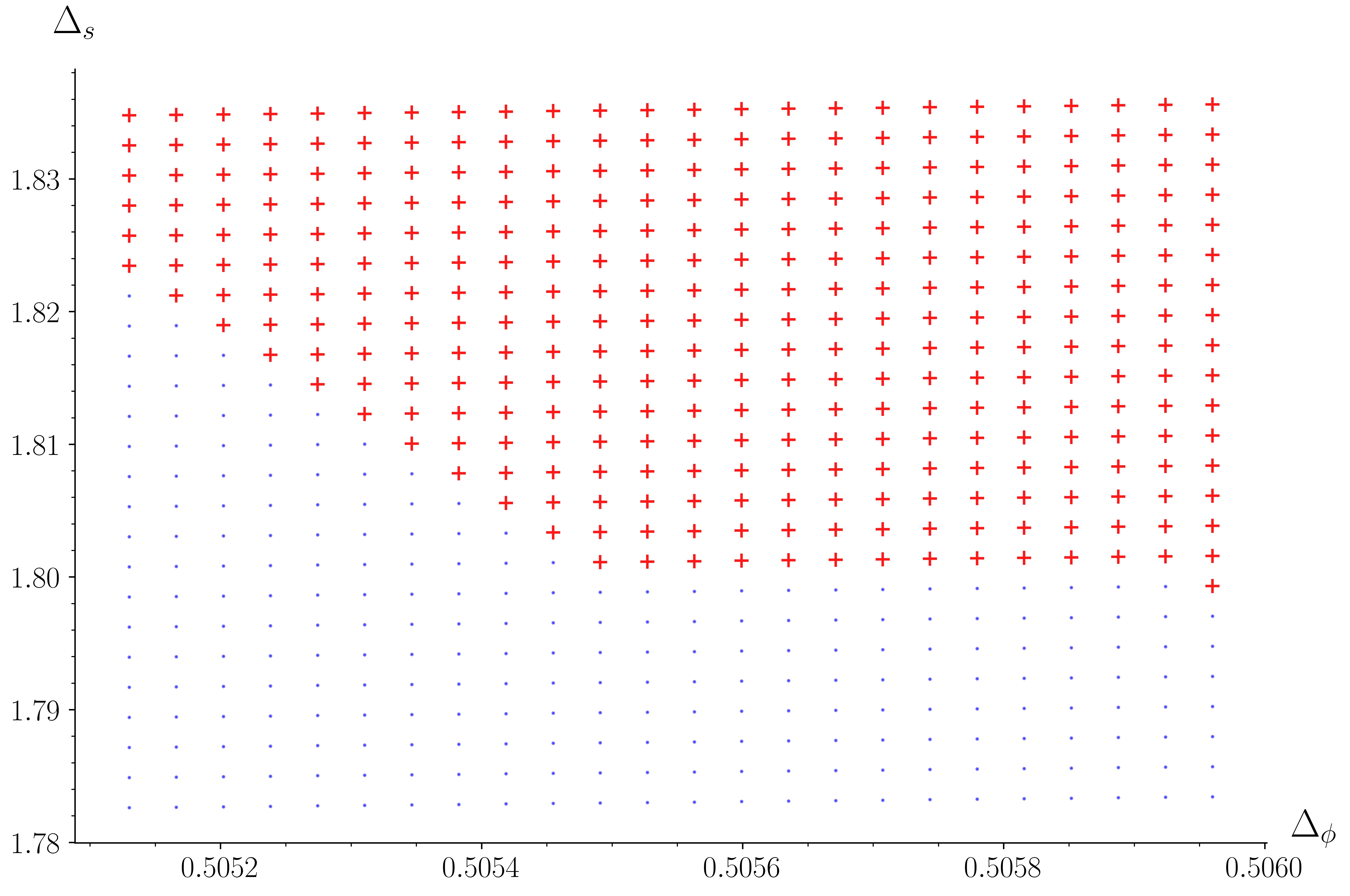}
\caption{\label{fig:8} Mixed-correlator conformal bootstrap results for derivative order $n_{\rm max} = 8$, under the same assumptions as in figure \ref{fig:1}, for a small region around the sharp kink in the boundary of the `allowed' region.  Red crosses mark the points that were determined to be primal feasible, i.e.\ `allowed'; blue dots mark the points that were determined to be dual feasible, i.e.\ disallowed.  Even at this resolution, the kink appears sharp, suggesting that there is a critical theory at $(\Delta_\phi,\Delta_s) \approx (0.5055,1.802)$, despite the lack of any large-$N$ prediction of such a critical point.}
\end{figure}

Finally, in figure \ref{fig:8} we show a close-up of the region where we saw a possible kink in the boundary of the `allowed' region.  We see that the kink in the boundary remains sharp even at this high resolution. Bootstrap phenomenology would suggest that such a sharp kink corresponds to a fixed point, leading us to tentatively conjecture the existence of a hitherto unknown $d=3$, $O(15) \otimes O(3)$ CFT that satisfies our spectral assumptions with scaling dimensions $\left(\Delta_{\phi}, \Delta_{s}\right)\approx\left(0.5055, 1.802\right)$.  If this additional CFT does exist, it may account for the fact that the large-$N$-predicted location of the chiral fixed point does not lie quite at the boundary of our `allowed' region:\ the upward advance of the disallowed region gets `stuck' on this new fixed point before it reaches the chiral one.  This, however, is speculation:\ further work, which we discuss briefly below, will be needed to clarify the situation.

\begin{figure}[tp]
\centering
\includegraphics[width=0.75\textwidth]{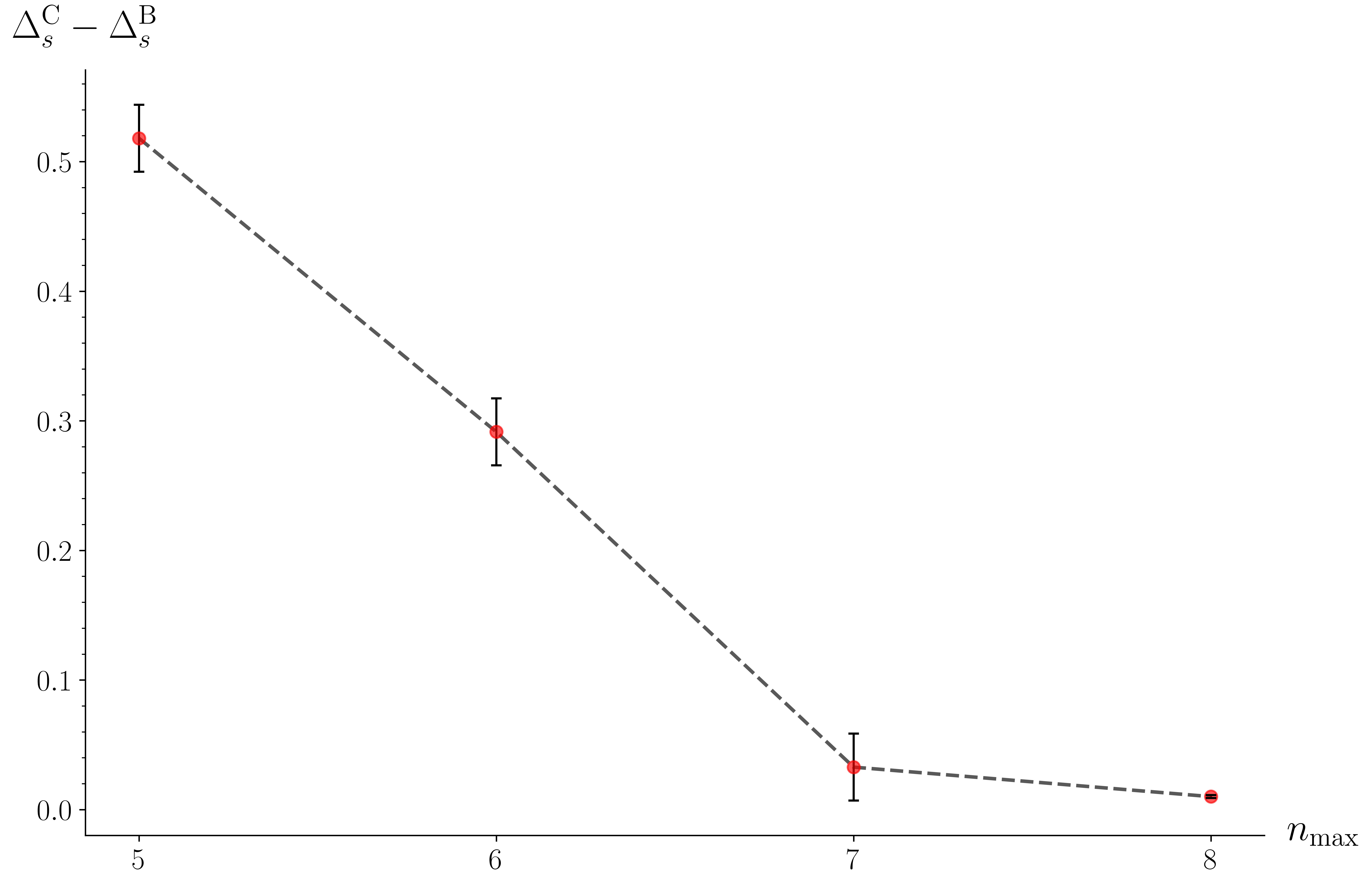}
\caption{\label{fig:9} The distance between the scaling dimension $\Delta_s$ at the lower boundary of the `allowed' region, $\Delta_s^{\rm B}$, and its large-$N$-predicted value at the chiral fixed point, $\Delta_s^{\rm C}$, as a function of the derivative order $n_{\rm max}$.  We estimate $\Delta_{s}^{\rm B}$ by taking it to lie half way between the last row of disallowed points and the first row of `allowed' points; the error bars depict the maximum possible error incurred via this method of estimation.}
\end{figure}

\section{Discussion} \label{s:discussion}
In this paper we have used the mixed-correlator conformal bootstrap technique to investigate the space of conformal field theories with $O(15) \otimes O(3)$ symmetry in $d=3$.  Our approach is to survey a two-dimensional projection of the space of scaling dimensions using as minimal a set of assumptions as possible, with the hope of thereby seeing signatures of more than one fixed point (e.g.\ Heisenberg and chiral) in the same calculation.

The essential features of our results are as follows:
\begin{itemize}
\item
We see pretty convincing features of the Heisenberg point --- see especially figures \ref{fig:5} and \ref{fig:6}.  There is a sharp kink in the boundary at the top left of the `allowed' region, and the location of this kink coincides very closely with the large-$N$ predictions for the scaling dimensions of $s$, the most relevant singlet operator, and $\phi$, the most relevant bifundamental operator, at that fixed point.
\item
We see only circumstantial evidence of the chiral point --- see especially figures \ref{fig:5} and \ref{fig:7}.  In figure \ref{fig:5} it looks as if the large-$N$-predicted location of the chiral point is actually on the boundary of the `allowed' region, but closer inspection (figure \ref{fig:7}) shows that it is actually still slightly inside it.  In neither case is there a sharp feature, e.g.\ a kink, in the boundary near the chiral point.  On the other hand, the upward advance of the disallowed region with increasing derivative order appears to have been halted at a level very close to that of the chiral point.  Figure \ref{fig:9} shows the distance between the large-$N$ prediction, $\Delta_{s}^{\rm C}$, and our bound as a function of the derivative order.  The fact that, within errors, the boundary does not advance between derivative order $n_{\rm max} = 7$ and $n_{\rm max}=8$ suggests that the stable chiral fixed point may be obstructing any further movement.
\item
Interestingly, we see what appears to be evidence of a third fixed point --- see especially figures \ref{fig:5} and \ref{fig:8}.  In both of these figures we see a sharp kink in the boundary of the `allowed' region at its lower left corner:\ a point that does not correspond to any of the fixed points predicted by the large-$N$ analyses in the literature.  If there is indeed a fixed point there, it might also account for why the lower edge of the `allowed' region stopped short of the chiral point:\ it got stuck (as it were) on this third fixed point before advancing that far.
\item
The predicted scaling dimensions of $s$ and $\phi$ at the antichiral point lie well within the disallowed region.  That is not a surprise, since the antichiral theory is expected to contain an additional relevant scalar singlet operator, meaning that it should not be `allowed' under the set of assumptions that we have used here.
\end{itemize}

One might argue that the simplest explanation of the apparent third fixed point, especially in view of the lack of direct signatures of the chiral point, is that the third fixed point {\it is\/} the chiral point.  For this to be true, however, the large-$N$ prediction of the anomalous dimension of $\phi$ at the chiral point would have to be off by a factor of 3, which seems unlikely when $N$ is as high as 15.  Furthermore, although they do not remark on it, Nakayama and Ohtsuki's single-correlator bootstrap calculations \cite{nakayama2014} also seem to show signatures of a third fixed point in addition to the chiral and antichiral ones.  In their figure 4, there definitely appear to be two kinks in the single-correlator bound:\ an upward kink at $\Delta_\phi \approx 0.5065$ and a downward one at $\Delta_\phi \approx 0.515$.  The latter is close to the predicted scaling dimension of $\phi$ at the antichiral point, while the former is close to the scaling dimension of $\phi$ at our unidentified third fixed point.  In their figure 5, one could argue that there are also two kinks, this time both downward:\ one at $\Delta_\phi \approx 0.505$ and a second at $\Delta_\phi \approx 0.517$.  The latter is close to the predicted scaling dimension of $\phi$ at the chiral point, while the former is close to the scaling dimension of $\phi$ at our unidentified third fixed point.

In future work, it would be natural to investigate whether the `allowed' region in the $O(15) \otimes O(3)$ case could be reduced further, and indeed whether it would eventually split into disconnected islands, one centred on each fixed point.  Such behaviour was seen in Kos {\it et al.\/}'s 2015 paper \cite{kos2015archipelago}; there, however, only one fixed point was visible in the $(\Delta_\phi,\Delta_s)$ plane, whereas here we expect two, or indeed three if the apparent third fixed point is real.  This could be attempted by increasing the derivative order $n_{\rm max}$, increasing the number of spins $\ell$, or both, though of course there are increases in computational resource associated with either of these steps.

It would also be good to find out more about the apparent third fixed point.  For example, one could undertake a sequence of conformal bootstrap studies on the groups $O(N) \otimes O(3)$ for various values of $N$, to see what happens to the kink as $N$ is varied.  Since it does not seem to be there in the large-$N$ theory, one possibility is that it merges with the Heisenberg point before the $N \to \infty$ limit is reached:\ this conjecture seems worthy of further investigation.

The fate of the third fixed point when $N$ is reduced would also be worth investigating.  In that connection, we note that there also appear to be more kinks in Nakayama and Ohtsuki's single-correlator bootstrap results \cite{nakayama2015} for $O(4) \otimes O(2)$ and $O(3) \otimes O(2)$ than they explicitly identify.  Specifically, we note an additional upward kink in their figure 3 at $\Delta_\phi \approx 0.516$, and possibly also an additional upward kink in their figure 2 at $\Delta_\phi \approx 0.521$.  If these also represent additional fixed points, it would be very interesting to see whether they have more clearly visible signatures in a mixed-correlator bootstrap treatment.

Finally, it would of course be possible to use the results of the semidefinite programme solver to give information about the spectrum of the conformal field theory to which the apparent third fixed point corresponds.  Techniques for such spectrum extraction can be found in the conformal bootstrap literature \cite{elshowk2013,elshowk2014,komargodski2017,simmonsduffin2017}.  However, applying them to this case would require further computation, since (due to storage restrictions) we were unable to keep a record of the solutions to the semidefinite programmes as we went along.  We therefore reserve this interesting topic for a future publication.

\appendix
\section{Derivation of bootstrap equations} \label{app:bootstrapeqns}
We give here the precise details of the derivation of the thirteen bootstrap equations (or `sum rules') that we use in this paper to place bounds on the scaling dimensions of $SS$ and $VV$ operators in $O(N) \otimes O(M)$ CFTs.  As mentioned in the main text, these are derived by equating the relevant conformal block decomposition (\ref{decomp}) in the 12-channel (i.e.\ the channel in which the first operator is paired with the second) with the decomposition in the 14-channel.

For the $G^{(VV)^4}$ correlator, the 12-channel decomposition yields
\bea
& \ds x_{12}^{2\Delta_\phi} x_{34}^{2\Delta_\phi} \left\langle \phi_{i\alpha}(\vec{x}_1) \phi_{j\beta}(\vec{x}_2) \phi_{k\gamma}(\vec{x}_3) \phi_{l\delta}(\vec{x}_4) \right\rangle = \sum_{{SS}^{+}} \lambda_{\phi\phi\co}^2 \chi^S_{ijkl} \chi^S_{\alpha\beta\gamma\delta} g_{\Delta,\ell}(u,v) \qquad \qquad & \nonumber \\
& \ds \qquad \qquad \qquad \qquad +\,\sum_{{ST}^{+}} \lambda_{\phi\phi\co}^2 \chi^S_{ijkl} \chi^{T,M}_{\alpha\beta\gamma\delta} g_{\Delta,\ell}(u,v) - \sum_{{SA}^{-}} \lambda_{\phi\phi\co}^2 \chi^S_{ijkl} \chi^A_{\alpha\beta\gamma\delta} g_{\Delta,\ell}(u,v) \qquad
& \nonumber \\
& \ds \qquad \qquad \qquad \qquad +\,\sum_{{TS}^{+}} \lambda_{\phi\phi\co}^2 \chi^{T,N}_{ijkl} \chi^S_{\alpha\beta\gamma\delta} g_{\Delta,\ell}(u,v) + \sum_{{TT}^{+}} \lambda_{\phi\phi\co}^2 \chi^{T,N}_{ijkl} \chi^{T,M}_{\alpha\beta\gamma\delta} g_{\Delta,\ell}(u,v) \qquad & \nonumber \\
& \ds \qquad \qquad \qquad \qquad -\,\sum_{{TA}^{-}} \lambda_{\phi\phi\co}^2 \chi^{T,N}_{ijkl} \chi^A_{\alpha\beta\gamma\delta} g_{\Delta,\ell}(u,v) - \sum_{{AS}^{-}} \lambda_{\phi\phi\co}^2 \chi^A_{ijkl} \chi^S_{\alpha\beta\gamma\delta} g_{\Delta,\ell}(u,v) \qquad & \nonumber \\
& \ds \qquad \qquad \qquad \qquad -\,\sum_{{AT}^{-}} \lambda_{\phi\phi\co}^2 \chi^A_{ijkl} \chi^{T,M}_{\alpha\beta\gamma\delta} g_{\Delta,\ell}(u,v) + \sum_{{AA}^{+}} \lambda_{\phi\phi\co}^2 \chi^A_{ijkl} \chi^A_{\alpha\beta\gamma\delta} g_{\Delta,\ell}(u,v). \qquad &
\eea
Here $\chi$ denotes the four-point structure that arises from contracting the vector indices of the $\phi$ operators in each different representation:
\bea
\chi^S_{ijkl} & = & \delta_{ij} \delta_{kl}; \\
\chi^{T,N}_{ijkl} & = & \delta_{il} \delta_{jk} + \delta_{ik} \delta_{jl} - \frac{2}{N} \delta_{ij} \delta_{kl}; \\
\chi^A_{ijkl} & = & \delta_{il} \delta_{jk} - \delta_{ik} \delta_{jl}.
\eea
Notice that only one of these three, $\chi^T$, depends explicitly on $N$.
The 14-channel decomposition is obtained by making the exchanges $j \leftrightarrow l$, $\beta \leftrightarrow \delta$, and $\vec{x}_2 \leftrightarrow \vec{x}_4$.  The result is
\bea
& \ds x_{14}^{2\Delta_\phi} x_{23}^{2\Delta_\phi} \left\langle \phi_{i\alpha}(\vec{x}_1) \phi_{j\beta}(\vec{x}_2) \phi_{k\gamma}(\vec{x}_3) \phi_{l\delta}(\vec{x}_4) \right\rangle = \sum_{{SS}^{+}} \lambda_{\phi\phi\co}^2 \chi^S_{ilkj} \chi^S_{\alpha\delta\gamma\beta} g_{\Delta,\ell}(v,u) \qquad \qquad & \nonumber \\
& \ds \qquad \qquad \qquad \qquad +\,\sum_{{ST}^{+}} \lambda_{\phi\phi\co}^2 \chi^S_{ilkj} \chi^{T,M}_{\alpha\delta\gamma\beta} g_{\Delta,\ell}(v,u) - \sum_{{SA}^{-}} \lambda_{\phi\phi\co}^2 \chi^S_{ilkj} \chi^A_{\alpha\delta\gamma\beta} g_{\Delta,\ell}(v,u) \qquad & \nonumber \\
& \ds \qquad \qquad \qquad \qquad +\,\sum_{{TS}^{+}} \lambda_{\phi\phi\co}^2 \chi^{T,N}_{ilkj} \chi^S_{\alpha\delta\gamma\beta} g_{\Delta,\ell}(v,u) + \sum_{{TT}^{+}} \lambda_{\phi\phi\co}^2 \chi^{T,N}_{ilkj} \chi^{T,M}_{\alpha\delta\gamma\beta} g_{\Delta,\ell}(v,u) \qquad & \nonumber \\
& \ds \qquad \qquad \qquad \qquad -\,\sum_{{TA}^{-}} \lambda_{\phi\phi\co}^2 \chi^{T,N}_{ilkj} \chi^A_{\alpha\delta\gamma\beta} g_{\Delta,\ell}(v,u) - \sum_{{AS}^{-}} \lambda_{\phi\phi\co}^2 \chi^A_{ilkj} \chi^S_{\alpha\delta\gamma\beta} g_{\Delta,\ell}(v,u) \qquad & \nonumber \\
& \ds \qquad \qquad \qquad \qquad -\,\sum_{{AT}^{-}} \lambda_{\phi\phi\co}^2 \chi^A_{ilkj} \chi^{T,M}_{\alpha\delta\gamma\beta} g_{\Delta,\ell}(v,u) + \sum_{{AA}^{+}} \lambda_{\phi\phi\co}^2 \chi^A_{ilkj} \chi^A_{\alpha\delta\gamma\beta} g_{\Delta,\ell}(v,u), \qquad & 
\eea
where we have used the fact, evident from (\ref{crossratios}), that $\vec{x}_2 \leftrightarrow \vec{x}_4$ implies $u \leftrightarrow v$.
Equating these two expressions for the $G^{(VV)^4}$ correlator, and separating the coefficients of each of the nine different tensor structures $\delta_{ij} \delta_{kl} \delta_{\alpha\beta} \delta_{\gamma\delta}$ etc.,\ we obtain nine equations.  Symmetrising and antisymmetrising these under the exchange $u \leftrightarrow v$, and defining \\
\be
\ds F_{\pm,\Delta,\ell}^{pq,rt}(u,v) \equiv v^{\frac{\Delta_{r}+\Delta_{q}}{2}}g_{\Delta,\ell}^{\Delta_{pq},\Delta_{rt}}(u,v) \pm u^{\frac{\Delta_{r}+\Delta_{q}}{2}}g_{\Delta,\ell}^{\Delta_{pq},\Delta_{rt}}(v,u),
\ee
we find
\bea
\sum_{SS^{+}} \ls F_{+,\Delta,\ell}^{\phi\phi,\phi\phi}(u,v) - \frac{2}{M}\sum_{ST^{+}}\ls F_{+,\Delta,\ell}^{\phi\phi,\phi\phi}(u,v) - \frac{2}{N}\sum_{TS^{+}}\ls F_{+,\Delta,\ell}^{\phi\phi,\phi\phi}(u,v) \nonumber\\
- \left(1-\frac{4}{NM}\right)\sum_{TT^{+}}\ls F_{+,\Delta,\ell}^{\phi\phi,\phi\phi}(u,v) + \sum_{TA^{-}}\ls F_{+,\Delta,\ell}^{\phi\phi,\phi\phi}(u,v)  \nonumber\\
+ \sum_{AT^{-}}\ls F_{+,\Delta,\ell}^{\phi\phi,\phi\phi}(u,v) - \sum_{AA^{+}}\ls F_{+,\Delta,\ell}^{\phi\phi,\phi\phi}(u,v) = 0, \label{vvvv1}
\eea
\bea
\sum_{SS^{+}} \ls F_{-,\Delta,\ell}^{\phi\phi,\phi\phi}(u,v) - \frac{2}{M}\sum_{ST^{+}}\ls F_{-,\Delta,\ell}^{\phi\phi,\phi\phi}(u,v) - \frac{2}{N}\sum_{TS^{+}}\ls F_{-,\Delta,\ell}^{\phi\phi,\phi\phi}(u,v) \nonumber\\ 
+ \left(1+\frac{4}{NM}\right)\sum_{TT^{+}}\ls F_{-,\Delta,\ell}^{\phi\phi,\phi\phi}(u,v) - \sum_{TA^{-}}\ls F_{-,\Delta,\ell}^{\phi\phi,\phi\phi}(u,v) \nonumber\\
- \sum_{AT^{-}}\ls F_{-,\Delta,\ell}^{\phi\phi,\phi\phi}(u,v) + \sum_{AA^{+}}\ls F_{-,\Delta,\ell}^{\phi\phi,\phi\phi}(u,v) = 0, \label{vvvv2}
\eea
\bea
\sum_{ST^{+}}\ls F_{+,\Delta,\ell}^{\phi\phi,\phi\phi}(u,v) + \sum_{SA^{-}}\ls F_{+,\Delta,\ell}^{\phi\phi,\phi\phi}(u,v) - \left(1+\frac{2}{N}\right)\sum_{TT^{+}}\ls F_{+,\Delta,\ell}^{\phi\phi,\phi\phi}(u,v) \nonumber\\
- \left(1+\frac{2}{N}\right)\sum_{TA^{-}}\ls F_{+,\Delta,\ell}^{\phi\phi,\phi\phi}(u,v) +\sum_{AT^{-}}\ls F_{+,\Delta,\ell}^{\phi\phi,\phi\phi}(u,v) + \sum_{AA^{+}}\ls F_{+,\Delta,\ell}^{\phi\phi,\phi\phi}(u,v) = 0, \nonumber \\
\label{vvvv3}
\eea
\bea
\sum_{ST^{+}}\ls F_{-,\Delta,\ell}^{\phi\phi,\phi\phi}(u,v) + \sum_{SA^{-}}\ls F_{-,\Delta,\ell}^{\phi\phi,\phi\phi}(u,v) + \left(1-\frac{2}{N}\right)\sum_{TT^{+}}\ls F_{-,\Delta,\ell}^{\phi\phi,\phi\phi}(u,v) \nonumber\\
+ \left(1-\frac{2}{N}\right)\sum_{TA^{-}}\ls F_{-,\Delta,\ell}^{\phi\phi,\phi\phi}(u,v) - \sum_{AT^{-}}\ls F_{-,\Delta,\ell}^{\phi\phi,\phi\phi}(u,v) - \sum_{AA^{+}}\ls F_{-,\Delta,\ell}^{\phi\phi,\phi\phi}(u,v) = 0, \nonumber \\
 \label{vvvv4}
\eea
\bea
\sum_{ST^{+}}\ls F_{+,\Delta,\ell}^{\phi\phi,\phi\phi}(u,v) - \sum_{SA^{-}}\ls F_{+,\Delta,\ell}^{\phi\phi,\phi\phi}(u,v) - \sum_{TS^{+}}\ls F_{+,\Delta,\ell}^{\phi\phi,\phi\phi}(u,v) \nonumber\\ + \left(\frac{2}{M}-\frac{2}{N}\right)\sum_{TT^{+}}\ls F_{+,\Delta,\ell}^{\phi\phi,\phi\phi}(u,v) + \frac{2}{N}\sum_{TA^{-}}\ls F_{+,\Delta,\ell}^{\phi\phi,\phi\phi}(u,v) \nonumber\\ + \sum_{AS^{-}}\ls F_{+,\Delta,\ell}^{\phi\phi,\phi\phi}(u,v) - \frac{2}{M}\sum_{AT^{-}}\ls F_{+,\Delta,\ell}^{\phi\phi,\phi\phi}(u,v) = 0, \label{vvvv5}
\eea
\bea
\sum_{ST^{+}}\ls F_{-,\Delta,\ell}^{\phi\phi,\phi\phi}(u,v) - \sum_{SA^{-}}\ls F_{-,\Delta,\ell}^{\phi\phi,\phi\phi}(u,v) + \sum_{TS^{+}}\ls F_{-,\Delta,\ell}^{\phi\phi,\phi\phi}(u,v) \nonumber\\ - \left(\frac{2}{M}+\frac{2}{N}\right)\sum_{TT^{+}}\ls F_{-,\Delta,\ell}^{\phi\phi,\phi\phi}(u,v) + \frac{2}{N}\sum_{TA^{-}}\ls F_{-,\Delta,\ell}^{\phi\phi,\phi\phi}(u,v) \nonumber\\ - \sum_{AS^{-}}\ls F_{-,\Delta,\ell}^{\phi\phi,\phi\phi}(u,v) + \frac{2}{M}\sum_{AT^{-}}\ls F_{-,\Delta,\ell}^{\phi\phi,\phi\phi}(u,v) = 0, \label{vvvv6}
\eea
\bea
\sum_{TS^{+}}\ls F_{+,\Delta,\ell}^{\phi\phi,\phi\phi}(u,v) - \left(1+\frac{2}{M}\right)\sum_{TT^{+}}\ls F_{+,\Delta,\ell}^{\phi\phi,\phi\phi}(u,v) + \sum_{TA^{-}}\ls F_{+,\Delta,\ell}^{\phi\phi,\phi\phi}(u,v) \nonumber\\
+\sum_{AS^{-}}\ls F_{+,\Delta,\ell}^{\phi\phi,\phi\phi}(u,v) - \left(1+\frac{2}{M}\right)\sum_{AT^{-}}\ls F_{+,\Delta,\ell}^{\phi\phi,\phi\phi}(u,v) + \sum_{AA^{+}}\ls F_{+,\Delta,\ell}^{\phi\phi,\phi\phi}(u,v) = 0, \nonumber \\
\label{vvvv7}
\eea
\bea
\sum_{TS^{+}}\ls F_{-,\Delta,\ell}^{\phi\phi,\phi\phi}(u,v) + \left(1-\frac{2}{M}\right)\sum_{TT^{+}}\ls F_{-,\Delta,\ell}^{\phi\phi,\phi\phi}(u,v) - \sum_{TA^{-}}\ls F_{-,\Delta,\ell}^{\phi\phi,\phi\phi}(u,v) \nonumber \\
+ \sum_{AS^{-}}\ls F_{-,\Delta,\ell}^{\phi\phi,\phi\phi}(u,v) + \left(1-\frac{2}{M}\right)\sum_{AT^{-}}\ls F_{-,\Delta,\ell}^{\phi\phi,\phi\phi}(u,v) - \sum_{AA^{+}}\ls F_{-,\Delta,\ell}^{\phi\phi,\phi\phi}(u,v) = 0, \nonumber \\
\label{vvvv8}
\eea
\bea
\sum_{TT^{+}}\ls F_{-,\Delta,\ell}^{\phi\phi,\phi\phi}(u,v) + \sum_{TA^{-}}\ls F_{-,\Delta,\ell}^{\phi\phi,\phi\phi}(u,v) \nonumber \\ + \sum_{AT^{-}}\ls F_{-,\Delta,\ell}^{\phi\phi,\phi\phi}u,v) + \sum_{AA^{+}}\ls F_{-,\Delta,\ell}^{\phi\phi,\phi\phi}(u,v) = 0. \label{vvvv9}
\eea
For the $G^{(SS)^4}$ correlator, the OPE decomposition in the 12-channel is
\be
\left\langle s(\vec{x}_1) s(\vec{x}_2) s(\vec{x}_3) s(\vec{x}_4) \right\rangle = x_{12}^{-2\Delta_s} x_{34}^{-2\Delta_s} \sum_{{SS}^{+}} \lambda_{ss\co}^2 g_{\Delta,\ell}(u,v),
\ee
while that in the 14-channel is
\be
\left\langle s(\vec{x}_1) s(\vec{x}_2) s(\vec{x}_3) s(\vec{x}_4) \right\rangle = x_{14}^{-2\Delta_s} x_{23}^{-2\Delta_s} \sum_{{SS}^{+}} \lambda_{ss\co}^2 g_{\Delta,\ell}(v,u).
\ee
Equating these yields
\be
\sum_{{SS}^{+}} \lambda_{ss\co}^2 F^{ss,ss}_{-,\Delta,\ell}(u,v)  = 0.
\ee
For the $G^{(VV)^2 (SS)^2}$ correlator, the 12-channel decomposition is
\be
\left\langle \phi_{i\alpha}(\vec{x}_1) \phi_{j\beta}(\vec{x}_2) s(\vec{x}_3) s(\vec{x}_4) \right\rangle = x_{12}^{-2\Delta_\phi} x_{34}^{-2\Delta_s} \sum_{{SS}^{+}} \lambda_{\phi\phi\co} \lambda_{ss\co} g_{\Delta,\ell}(u,v),
\ee
while the OPE decomposition in the 14-channel is
\bea
& \ds \left\langle \phi_{i\alpha}(\vec{x}_1) \phi_{j\beta}(\vec{x}_2) s(\vec{x}_3) s(\vec{x}_4) \right\rangle = \qquad \qquad \qquad \qquad \qquad \qquad \qquad \qquad \qquad \qquad \qquad \qquad & \nonumber \\
& \ds \qquad \left( x_{23} x_{14} \right)^{-\Delta_\phi-\Delta_s} \left( \frac{x_{24}}{x_{34}} \right)^{\Delta_{s\phi}} \left( \frac{x_{34}}{x_{13}} \right)^{\Delta_{\phi s}} \sum_{{VV}^{\pm}} \lambda_{s \phi \co} \lambda_{\phi s \co} \left( -1 \right)^\ell g^{\Delta_{s \phi},\Delta_{\phi s}}_{\Delta,\ell}(v,u), \qquad &
\eea
or, using the identity $\lambda_{s\phi\co} = (-1)^\ell \lambda_{\phi s \co}$,
\bea
& \ds \left\langle \phi_{i\alpha}(\vec{x}_1) \phi_{j\beta}(\vec{x}_2) s(\vec{x}_3) s(\vec{x}_4) \right\rangle = \qquad \qquad \qquad \qquad \qquad \qquad \qquad \qquad \qquad \qquad \qquad \qquad  & \nonumber \\
& \ds \qquad \left( x_{23} x_{14} \right)^{-\Delta_\phi-\Delta_s} \left( \frac{x_{24}}{x_{34}} \right)^{\Delta_{s\phi}} \left( \frac{x_{34}}{x_{13}} \right)^{\Delta_{\phi s}} \sum_{{VV}^{\pm}} \lambda_{\phi s \co}^2 g^{\Delta_{s \phi},\Delta_{\phi s}}_{\Delta,\ell}(v,u). \quad &
\eea
Equating the 12-channel and 14-channel decompositions yields the symmetrised and antisymmetrised equations
\be
\sum_{{SS}^{+}}\lambda_{\phi\phi\co}\lambda_{ss\co}F^{\phi\phi,ss}_{+,\Delta,\ell} - \sum_{{VV}^{+}}\lambda_{\phi{s}\co}^{2}F^{s \phi,\phi s}_{+,\Delta,\ell} - \sum_{{VV}^{-}}\lambda_{\phi{s}\co}^{2}F^{s \phi,\phi s}_{+,\Delta,\ell} = 0,
\ee
\be
\sum_{{SS}^{+}}\lambda_{\phi\phi\co}\lambda_{ss\co}F^{\phi\phi,ss}_{-,\Delta,\ell} + \sum_{{VV}^{+}}\lambda_{\phi{s}\co}^{2}F^{s \phi,\phi s}_{-,\Delta,\ell} + \sum_{{VV}^{-}}\lambda_{\phi{s}\co}^{2}F^{s \phi,\phi s}_{-,\Delta,\ell} = 0.
\ee
Finally, for the $G^{(VV)(SS)(VV)(SS)}$ correlator, the OPE decomposition in the 12-channel is
\bea
& \ds \left<\phi_{i\alpha}(\vec{x}_{1})s(\vec{x}_{2})\phi_{j\beta}(\vec{x}_{3})s(\vec{x}_{4})\right> = \qquad \qquad \qquad \qquad \qquad \qquad \qquad \qquad \qquad \qquad \qquad \qquad  & \nonumber \\
& \ds x_{12}^{-\left(\Delta_{\phi}+\Delta_{s}\right)} x_{34}^{-\left(\Delta_{\phi}+\Delta_{s}\right)} \left(\frac{x_{24}}{x_{14}}\right)^{\Delta_{\phi{s}}} \left(\frac{x_{14}}{x_{13}}\right)^{\Delta_{\phi{s}}} \sum_{VV^{\pm}}\lambda_{\phi{s}\co}^{2} (-1)^\ell g_{\Delta, \ell}^{\Delta_{\phi{s}},\Delta_{\phi{s}}}(u,v), \quad &
\eea
while that in the 14-channel is
\bea
& \ds \left<\phi_{i\alpha}(\vec{x}_{1})s(\vec{x}_{2})\phi_{j\beta}(\vec{x}_{3})s(\vec{x}_{4})\right> = \qquad \qquad \qquad \qquad \qquad \qquad \qquad \qquad \qquad \qquad \qquad \qquad & \nonumber \\
& \ds x_{23}^{-\left(\Delta_{\phi}+\Delta_{s}\right)} x_{14}^{-\left(\Delta_{\phi}+\Delta_{s}\right)} \left(\frac{x_{24}}{x_{34}}\right)^{\Delta_{\phi{s}}} \left(\frac{x_{34}}{x_{13}}\right)^{\Delta_{\phi{s}}} \sum_{VV^{\pm}}\lambda_{\phi{s}\co}^{2} (-1)^\ell g_{\Delta, \ell}^{\Delta_{\phi{s}},\Delta_{\phi{s}}}(v,u). \quad &
\eea
Equating these yields
\be
\sum_{{VV}^{\pm}}\lambda_{\phi s\co}^{2} (-1)^\ell F_{-,\Delta,\ell}^{\phi s, \phi s}(u,v) = 0,
\ee
or, separating the $VV{+}$ and $VV{-}$ sectors explicitly,
\be
\sum_{{VV}^{+}} \lambda_{\phi s\co}^{2} F_{-,\Delta,\ell}^{\phi s, \phi s}(u,v) 
- \sum_{{VV}^{-}} \lambda_{\phi s\co}^{2} F_{-,\Delta,\ell}^{\phi s, \phi s}(u,v) = 0.
\ee

All of these constraints can be encoded in a single 13-dimensional vectorial sum rule, where $\vec{V}_{SS,\Delta,\ell}$ is a 13-vector of $2 \times 2$ matrices and all the other ${\vec V}_{XY,\Delta,\ell}$ are 13-vectors of $1 \times 1$ matrices, i.e.\ scalars:
\bea
0 = \sum_{{SS}^{+}}
\begin{pmatrix}
\lambda_{\phi\phi\co} & \lambda_{ss\co}
\end{pmatrix}
\vec{V}_{SS,\Delta, \ell}
\begin{pmatrix}
\lambda_{\phi\phi\co} \\
\lambda_{ss\co}
\end{pmatrix}
 + \sum_{{ST}^{+}}\lambda_{\phi\phi\co}^{2}\vec{V}_{ST,\Delta, \ell} + \sum_{{SA}^{-}}\lambda_{\phi\phi\co}^{2}\vec{V}_{SA,\Delta, \ell} \nonumber \\
+ \sum_{{TS}^{+}}\lambda_{\phi\phi\co}^{2}\vec{V}_{TS,\Delta, \ell} + \sum_{{TT}^{+}}\lambda_{\phi\phi\co}^{2}\vec{V}_{TT,\Delta, \ell} + \sum_{{TA}^{-}}\lambda_{\phi\phi\co}^{2}\vec{V}_{TA,\Delta, \ell} \nonumber \\
+ \sum_{{AS}^{-}}\lambda_{\phi\phi\co}^{2}\vec{V}_{AS,\Delta, \ell} + \sum_{{AT}^{-}}\lambda_{\phi\phi\co}^{2}\vec{V}_{AT,\Delta, \ell} + \sum_{{AA}^{+}}\lambda_{\phi\phi\co}^{2}\vec{V}_{AA,\Delta, \ell} \nonumber \\
+ \sum_{{VV}^{+}}\lambda_{\phi{s}\co}^{2}\vec{V}_{VV+,\Delta, \ell} + \sum_{{VV}^{-}}\lambda_{\phi{s}\co}^{2}\vec{V}_{VV-,\Delta, \ell},
\eea
where
\small
\begin{align*}
\vec{V}_{SS, \Delta, \ell} = \left(
\begin{array}{c}
\begin{pmatrix}
\fp & 0 \\
0 & 0
\end{pmatrix}\\
\begin{pmatrix}
\fm & 0 \\
0 & 0
\end{pmatrix}\\
\begin{pmatrix}
0 & 0 \\
0 & 0
\end{pmatrix}\\
\begin{pmatrix}
0 & 0 \\
0 & 0
\end{pmatrix}\\
\begin{pmatrix}
0 & 0 \\
0 & 0
\end{pmatrix}\\
\begin{pmatrix}
0 & 0 \\
0 & 0
\end{pmatrix}\\
\begin{pmatrix}
0 & 0 \\
0 & 0
\end{pmatrix}\\
\begin{pmatrix}
0 & 0 \\
0 & 0
\end{pmatrix}\\
\begin{pmatrix}
0 & 0 \\
0 & 0
\end{pmatrix}\\
\begin{pmatrix}
0 & 0 \\
0 & F^{ss,ss}_{-,\Delta,\ell}
\end{pmatrix}\\
\begin{pmatrix}
0 & \frac{1}{2}F^{\phi\phi,ss}_{+,\Delta,\ell} \\
\frac{1}{2}F^{\phi\phi,ss}_{+,\Delta,\ell} & 0
\end{pmatrix}\\
\begin{pmatrix}
0 & \frac{1}{2}F^{\phi\phi,ss}_{-,\Delta,\ell} \\
\frac{1}{2}F^{\phi\phi,ss}_{-,\Delta,\ell} & 0
\end{pmatrix}\\
\begin{pmatrix}
0 & 0 \\
0 & 0
\end{pmatrix}\\
\end{array}
\right), 
\vec{V}_{ST, \Delta, \ell} = \left(
\begin{array}{c}
-\frac{2}{M}\fp\\
-\frac{2}{M}\fm\\
\fp\\
\fm\\
\fp\\
\fm\\
0\\
0\\
0\\
0\\
0\\
0\\
0\\
\end{array}
\right),
\vec{V}_{SA, \Delta, \ell} = \left(
\begin{array}{c}
0\\
0\\
\fp\\
\fm\\
-\fp\\
-\fm\\
0\\
0\\
0\\
0\\
0\\
0\\
0\\
\end{array}
\right),
\end{align*}
\begin{align*}
\vec{V}_{TS, \Delta, \ell}=\left(
\begin{array}{c}
-\frac{2}{N}\fp\\
-\frac{2}{N}\fm\\
0\\
0\\
-\fp\\
\fm\\
\fp\\
\fm\\
0\\
0\\
0\\
0\\
0\\
\end{array}
\right)\!\!,
\vec{V}_{TT, \Delta, \ell}=\left(
\begin{array}{c}
-\left(1-\frac{4}{NM}\right)\fp\\
\left(1+\frac{4}{NM}\right)\fm\\
-\left(1+\frac{2}{N}\right)\fp\\
\left(1-\frac{2}{N}\right)\fm\\
\left(\frac{2}{M}-\frac{2}{N}\right)\fp\\
-\left(\frac{2}{M}+\frac{2}{N}\right)\fm\\
-\left(1+\frac{2}{M}\right)\fp\\
\left(1-\frac{2}{M}\right)\fm\\
\fm\\
0\\
0\\
0\\
0\\
\end{array}
\right)\!\!,
\vec{V}_{TA, \Delta, \ell}=\left(
\begin{array}{c}
\fp\\
-\fm\\
-\left(1+\frac{2}{N}\right)\fp\\
\left(1-\frac{2}{N}\right)\fm\\
\frac{2}{N}\fp\\
\frac{2}{N}\fm\\
\fp\\
-\fm\\
\fm\\
0\\
0\\
0\\
0\\
\end{array}
\right),
\end{align*}
\begin{align*}
\vec{V}_{AS, \Delta, \ell} = \left(
\begin{array}{c}
0\\
0\\
0\\
0\\
\fp\\
-\fm\\
\fp\\
\fm\\
0\\
0\\
0\\
0\\
0\\
\end{array}
\right),
\vec{V}_{AT, \Delta, \ell} = \left(
\begin{array}{c}
\fp\\
-\fm\\
\fp\\
-\fm\\
-\frac{2}{M}\fp\\
\frac{2}{M}\fm\\
-\left(1+\frac{2}{M}\right)\fp\\
\left(1-\frac{2}{M}\right)\fm\\
\fm\\
0\\
0\\
0\\
0\\
\end{array}
\right),
\vec{V}_{AA, \Delta, \ell} = \left(
\begin{array}{c}
-\fp\\
\fm\\
\fp\\
-\fm\\
0\\
0\\
\fp\\
-\fm\\
\fm\\
0\\
0\\
0\\
0\\
\end{array}
\right),
\end{align*}
\be
\vec{V}_{VV+, \Delta, \ell} = \left(
\begin{array}{c}
0\\
0\\
0\\
0\\
0\\
0\\
0\\
0\\
0\\
0\\
-F^{s \phi,\phi s}_{+,\Delta,\ell}\\
F^{s \phi,\phi s}_{-,\Delta,\ell}\\
F^{\phi s,\phi s}_{-,\Delta,\ell}\\
\end{array}
\right),
\vec{V}_{VV-, \Delta, \ell} = \left(
\begin{array}{c}
0\\
0\\
0\\
0\\
0\\
0\\
0\\
0\\
0\\
0\\
-F^{s \phi,\phi s}_{+,\Delta,\ell}\\
F^{s \phi,\phi s}_{-,\Delta,\ell}\\
-F^{\phi s,\phi s}_{-,\Delta,\ell}\\
\end{array}
\right).
\ee
\normalsize
Note that we are using the convention used in \cite{kos2015archipelago}, which necessitates a factor of $(-1)^{\ell}$ in front of the conformal blocks compared to the convention in \cite{kos2014vector}.  The $(u,v)$-dependence of the convolved conformal blocks, $F_{\pm,\Delta,\ell}^{pq,rt}(u,v)$, has been suppressed in the vectors for clarity.
\acknowledgments
We are grateful to Connor Behan, Yu Nakayama, David Poland, Sam Ridgway, Slava Rychkov, and Alessandro Vichi for useful discussions.  This work used the ARCHER UK National Supercomputing Service (\url{http://www.archer.ac.uk}).  MTD and CAH acknowledge financial support from UKRI via EPSRC grant numbers EP/R513337/1 and EP/R031924/1 respectively.

\providecommand{\href}[2]{#2}\begingroup\raggedright\endgroup
\end{document}